\documentclass[a4paper, 10pt]{article}      
\usepackage[a4paper, total={6in, 8in}]{geometry}
\usepackage{graphicx}
\usepackage{amsmath}
\usepackage{bm}
\usepackage{amsfonts}
\usepackage{amssymb}
\usepackage{mathtools}
\usepackage{mathrsfs}
\usepackage{multirow}
\usepackage{url}
\usepackage{algorithm} 
\usepackage{algpseudocode}
\usepackage[draft]{fixme}
\usepackage{relsize}
\usepackage{authblk}
\usepackage{epstopdf}
\usepackage{color, colortbl}
\usepackage{hyperref}
\usepackage{cite}
\usepackage{caption}

\title{\LARGE \bf
Digital twins with distributed particle simulation for mine-to-mill material tracking}

\author{Martin Servin$^{1}$, Folke Vesterlund, and Erik Wallin$^{1}$
\thanks{$^{1}$Department of Physics, Ume\aa\ University, \{martin.servin, folke.vesterlund, erik.wallin\}@umu.se}%
\thanks{The project was funded in part by eSSENCE and VINNOVA (grant id 2019-04832).}
}%

\begin{document}

\maketitle
\thispagestyle{empty}
\pagestyle{empty}

\abstract{Systems for transport and processing of granular media are challenging to analyse, operate and optimise. In the mining and mineral processing industries these systems are chains of processes with complex interplay between the equipment, control, and the processed material. The material properties have natural variations that are usually only known at certain locations. Therefore, we explore a material-oriented approach to digital twins with a particle representation of the granular media. In digital form, the material is treated as pseudo-particles, each representing a large collection of real particles of various sizes, shapes and, mineral properties. Movements and changes in the state of the material are determined by the combined data from control systems, sensors, vehicle telematics, and simulation models at locations where no real sensors can see. The particle-based representation enables material tracking along the chain of processes. Each digital particle can act as a carrier of observational data generated by the equipment as it interacts with the real material. This makes it possible to better learn material properties from process observations, and to predict the effect on downstream processes. We test the technique on a mining simulator and demonstrate analysis that can be performed using data from cross-system material tracking.
}


\section{Introduction}

It is estimated that mining accounts for 6\% of the global energy consumption. Grinding is the most energy demanding process, accounting for 32\% of the mine's energy consumption, followed by haulage (24\%), underground mine ventilation (9\%), and digging (8\%) \cite{Holmberg2017}. 
The grinding efficiency is many times as low as 1\% and depends strongly on the particle size distribution and ore hardness \cite{Radziszewski2013}.
Therefore there is a strong motivation for determining and controlling the properties of the ore fed into the mill.
Mine-to-mill refers to a holistic approach to mine optimisation, see \cite{mckee:2013:umm} for a comprehensive description.
Instead of optimising each process individually, one takes the ore body characteristic into account, and consider the throughput and energy consumption of the full chain of processes when deciding on target fragmentation and controlling the feed to the mill.
The mine-to-mill approach has, however, proven difficult to implement in practice.
One reason is that it requires an elaborate digital infrastructure for collecting, processing and communicating data between edge devices, distributed control systems and centralized data services. The technology, popularized as digital twins \cite{grieves2014digital} or Industrial Internet of Things (IIoT) \cite{jeschke2017industrial}, is currently under intensive development.
A second challenge in implementing mine-to-mill is the unknown variations in the properties of the ore body.
The variations propagate through the process chain in a complicated way, and add to the full-system dynamics. 



\begin{figure}[h]	
\centering
  \captionsetup{width=0.85\textwidth}
\includegraphics[width=0.85\textwidth]{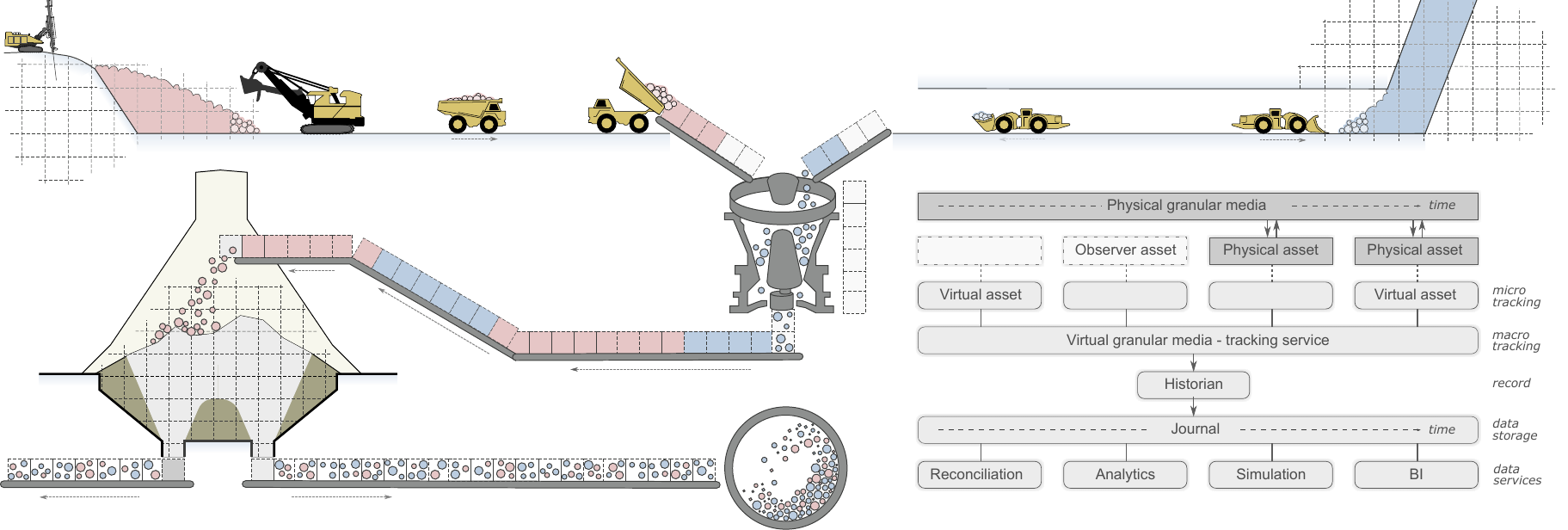}
\caption{Illustration of a mine-to-mill process with particle-based material tracking. The chain of processes include: drilling; blasting; load, haul and dump; conveying; crushing; sizing; storage and reclaim; and grinding; which then lead to a concentrator. 
The processes are carried out by physical assets that perform unit operations on the material.
Connected virtual assets mirror the operations by transforming a distributed digital representation of the material.
The architecture of the system of assets and services for tracking, data storage and analytics is illustrated in the lower right. 
}
\label{fig:particle_based_tracking}
\end{figure}  

A mine-to-mill process chain is illustrated in Fig.~\ref{fig:particle_based_tracking}.
The performance of each individual process depends on its design, control, and on the properties of the material feed. Each process also alters the state of the material. 
The complex interdependency between the process dynamics and variations in the material properties makes it hard to determine what is cause and effect.
The mineral concentration, hardness, strength, and toughness, of the ore body is partially known in advance, from exploratory drilling, and stored in a block model \cite{Rossi2014}.
These are input parameters to many models for blasting \cite{Ouchterlony2019}, crushing \cite{Evertsson2000} or grinding \cite{NapierMunn1996}, such as the Kuz–Ram blasting and JKMRC grinding models.
Measurement-while-drilling (MWD) of blast holes produce data that contain information about the geo-mechanical properties at higher spatial resolution than the exploration data. 
This may be used for improving the blast plans and for optimizing the subsequent processes \cite{Rai2016,Zhou2011a}.
Drilling and blasting is conducted in cycles, each cycle producing a volume of fragmented rock that is loaded using bucket excavators or load-haul-dump trucks before the next blasting.
The particle size and shape distribution is subsequently transformed by the crushers and grinders before entering the concentrator.
Variations in the diggability of the blasted rock \cite{Singh2006,Khorzoughi2016,Brunton2003} is useful for improving blast plans or for predicting the throughput at the crusher and mill.
Transport and storage systems alters the location of the material, but may also induce (unwanted) mixing and segregation that propagate to subsequent processes and affect the dynamics there.
Mine-to-mill analysis and optimization require knowledge of these changes but on-line measurements of granular media properties are inherently difficult.
The material movements during vehicle and belt conveyor transport are easily tracked and these systems may be instrumented with sensors for monitoring some of the material properties.
But in chutes, stockpiles and silos, little is known about the material beyond the knowledge of what is fed into them.
If any sensors can be installed at all, the observations are either indirect or limited to the surface of the granular media, which constitutes a small and perhaps not representative fraction of the bulk.
Conventional tracking systems, that rely on combinations of plug-flow models (or first-in/first-out) and perfect mixing models that are calibrated using tracer experiments \cite{kvarnstrom2012}, fall short on transport and storage systems with complex surface and discharge flow. 
This creates blind spots in the tracking of material movements and knowing the properties at any given location.

To remedy this, we explore digital twins with distributed particle simulation, of systems doing transport and processing of granular media. 
The idea is to represent the material and its movements in a structured data format based on a particle representation. 
In addition to the material's identity and current position, the particle data structure supports the reading and writing of observations from sensors and equipment along the chain of processes. 
When connected equipment performs unit operations on the material, the digital copy is updated accordingly.
When the material reaches blind spots in the system, for example a silo or a stockpile, the
digital copy is driven by a simulation model fed with data from the control system
and available sensors in real-time.
For the sake of computational performance and memory, the digital twin uses pseudo-particles that represents a large collection of real mineral particles with a distribution of grain size and shape.
For fast simulation we use the nonsmooth discrete element method, which allow large time-step integration \cite{Servin2014}.  
To achieve real-time performance in large processes, a data-driven model order reduction technique \cite{wallin:2021:ddm} is employed.
The combination of the two models in an executable application, that can be part of a distributed system like a digital twin, we call a \emph{granular surrogate}. 

We test the feasibility of doing material tracking with this technique on an idealized mine simulator with both discrete and continuous transport processes from two different sources. 
The tests include analysis of how crusher power draw and milling performance depends on variations in ore hardness and fragmentation at the blast location.
From crusher time-series it is possible to identify the relation between the ore hardness and measurement-while-drilling signal despite having multiple ore sources being mixed.
The sensitivity to tracking errors is analysed by perturbing the pseudo-particle origins.
The response in mill performance to a change in ore hardness at the source is tested with different state of an intermediate stockpile.
The effect is significant but it is still possible to reconcile the ore's grindability, in terms of the Bond Ball Mill index, to the locations of origin in the ore body thanks to the particle-based representation.

Related work include the data-driven mine-to-mill framework proposed and tested in \cite{Erkayaoglu2015} and \cite{Erkayaoglu2019}. 
It was concluded that material tracking needs to be performed in a more comprehensive way to better connect mining and mineral processing data, and stockpile management, 
and the required infrastructure was found to be the main limitation for implementing the framework. 
A closed-loop framework for real-time reserve management was introduced in \cite{Benndorf2016}.
It incorporate sensor-based material characterisation, geostatistical modeling under uncertainty, data assimilation for a sequential model updating, and mining system simulation and optimisation.
Using a modified Kalman filter technique, online sensor and process information is back-propagated into the resource block model, which in turn is used for production planning and other operative decisions.
Innes et al. \cite{innes:2011:ete,Innes2015} developed a method for representing, tracking and fusing information of excavated material.
The bulk material is represented as lumped masses, although not as particles and no other properties than mass and location.
Tracking along the transport and storage process is achieved using an Augmented State Kalman Filter with a mass preservation constraint. 
Reconciliation of the Bond Mill Work index to the ore block model using mine-to-mill tracking was demonstrated in \cite{Wambeke2018}, using a first-in-first-out model for the stockpile.
A 3D real-time stockpile model and mapping technique of product quality was developed in \cite{Zhao2016} for the purpose of planning and control in stacking, reclaiming and blending. 
The pile surface was reconstructed from the point-cloud of a mobile scanning device.
A voxelized model is then associated with a reclaiming machine to achieve autonomous operation and predictive quality.  
The simulator in \cite{Berton2013} demonstrate the combination of multiple, relatively simple, models for capturing the ore flow through a complex storage facility.
No previous work was found where material tracking is made using a particle representation.

\section{Digital twin as a distributed particle simulation}

From the material point of view, the digital twin of a mine may be considered a system of distributed particle simulation.
For computational efficiency, we use pseudo-particles, each representing a large collection of real rock particles of various size, shape and mineral composition.
Labelled $n$, a pseudo-particle have position $\bm{x}_n(t)\in\mathbb{R}^3$, velocity $\bm{v}_n(t) = \dot{\bm{x}}_n$,
mass $m_n \in\mathbb{R}^1$, size distribution function $\bm{f}_n \in\mathbb{R}^{N_{\text{f}}}$, mineral concentration $\bm{c}_n \in\mathbb{R}^{N_{\text{c}}}$, and mechanical properties $\bm{h}_n \in\mathbb{R}^{N_{\text{h}}}$ that characterize the real particles it represents, each discretized in classes of dimensionality $N_{\text{f}}$, $N_{\text{c}}$, and $N_{\text{h}}$, respectively.
The particle state vector is represented $\bm{X}_n =[m_n, \bm{x}_n, \bm{v}_n, \bm{f}_n, \bm{c}_n, \bm{h}_n]$.
Each particle has a position of origin, $\bm{x}_n(t_0)$.
The material properties may be known to some degree of certainty at that location from prior exploration and represented in a block model of grid cells indexed $i$, with centre point $\bm{x}_i$ and volume $V_i$. Or, they are the subject for determination by measurements in the process downstream.
The fragmentation is set by the blasting model and/or measurements after blasting.

\begin{figure}[H]
    \centering
  \captionsetup{width=0.85\textwidth}
  \includegraphics[width=0.85\textwidth]{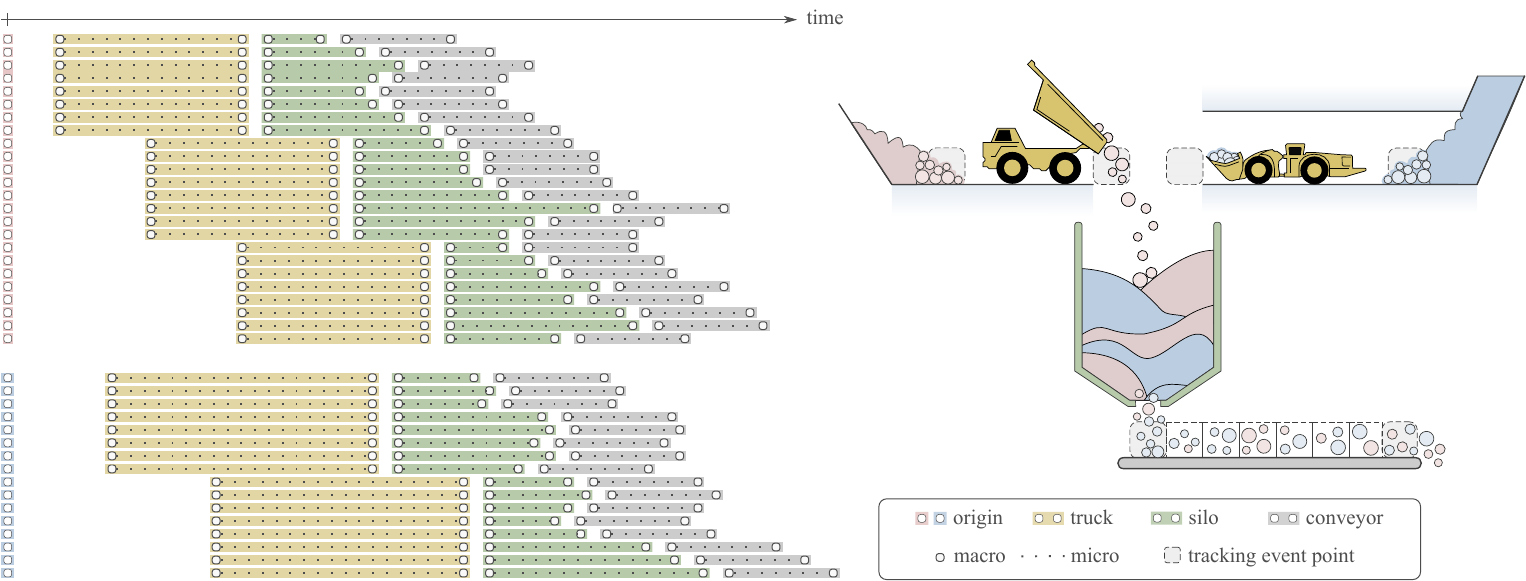}
    \caption{Illustration of micro and macro-tracking.
    The sparse macro-tracking events include the appearance of fragmented material at the source (pink or blue) after blasting, haul truck asset dispatching and dumping (yellow), material entering later being discharged from a storage asset (green) and a conveyor asset (grey).  
    The micro-tracking is the relatively high frequency tracking of the material moving with or flowing inside an asset.
    \label{fig:micro_macro_tracking}}
\end{figure}

The fixed and mobile mining equipment constitute a discrete set of \emph{assets} that transform the material by a set of unit operations, the primary two being \texttt{move} and \texttt{fragment}. 
Each pseudo-particle is tracked, or simulated, by a \emph{virtual asset} which is a realtime model that mirrors a physical asset and its unit operations on the material.
We distinguish between high-frequency \emph{micro-tacking} inside each asset and \emph{macro-tracking} of special events, e.g., when particles enter or exit an asset, or pass a point of special interest. This is illustrated in Fig.~\ref{fig:micro_macro_tracking}.

For a bucket excavator or truck asset, the micro-tracking can be accomplished simply by inheriting the position of the bucket or the truck during the transport from the loading to the dumping area.
Particles transported on a belt conveyor may similarly be tracked between the loading and discharge points by simply tracking the movement of the belt using the conveyor control system.
In the transporter's virtual asset, the belt is discretized longitudinally in bins of known location and particle content. 
Chutes, silos and stockpiles are different. 
In these assets the particle flow is passive, driven by gravity and feeders at the outlet for a controlled discharge rate.
Micro-tracking require a model, that may range from simple first-in-first-out or perfect mixing models, to a physics-based flow model.

Mathematically, we represent a unit operation by an asset $a$ as an operator $\hat{u}_a(t,\Delta t)$ that propagate a particle state $\bm{X}_n(t)$ forward in time into the new state $\bm{X}_n(t + \Delta t) = \hat{u}_a(t,\Delta t) \bm{X}_n(t)$, with propagation time-step $\Delta t$. 
If the particle $n$ enter the asset at time $t_\text{in}$ in a state $\bm{X}_n(t_\text{in})$, the exit state at time $t_\text{out}$ is given by 
\begin{equation}
    \bm{X}_n(t_\text{out}) = \prod_{j=0}^{J^a_n - 1} \hat{u}_a(t_\text{in} + j\Delta t){\bm{X}_n(t_\text{in})},
\end{equation}
where $J^a_n$ is the number of unit operations by the asset.
Micro-tracking is to resolve the evolution of the particles this way.
In general, the step size $\Delta t$ may vary and the unit operation is a function of both time and the state $\bm{X}_a$ of the particles in the asset , i.e., $\hat{u}_a(t,\bm{X}_a(t))$.
Note that this representation support both kinematic and dynamic models for the particle motion, for \texttt{move} operations.
Discrete element simulations, taking realtime asset data as input, is an example of the latter.
Letting the particles co-move with a vehicle on conveyor is an example of the former.
The evolution of the particle size distribution may also be represented by this, for \texttt{fragment} operations. 

Macro-tracking is represented as message passing, between virtual assets or a centralized tracking service, 
about events of special interest and the involved particles. 
The message should have a structured data format, like
\begin{align}
    \text{event} = & \bigl\{ \text{id}, \text{type}, \text{location}\ \bm{x}, \text{time}\ t, \text{particles} = \bigl\{ \left[n_1, \bm{X}_{n_1}(t) \right], \left[n_2, \bm{X}_{n_2}(t) \right], \hdots  \bigr\} \bigr\}
\end{align}
including an identifier for the event, its type, location and time, plus a set of particle identifiers and state tuples.
The macro-tracking data may then be stored in a database for later analytics or other services based on tracking data. An illustration of the architecture of the system of assets and services is found in Fig.~\ref{fig:particle_based_tracking}.

Most assets are equipped with sensors that produce observations while handling the material.
An observation is either a direct measurement of an intrinsic attribute or an indirect measurement.
The latter may be registered as an extrinsic attribute that is assigned to the pseudo-particles at the time and location for the observation.
Examples of extrinsic attributes are measurement-while-drilling (MWD) data, diggability, crushability, and grindability.
In complement to real sensors, it is also possible to produce \emph{virtual sensors} that enable monitoring or controlling a process at higher precision.
A virtual measurement signal of an attribute $p$ is accomplished by extracting the pseudo-particles $\mathcal{N}(\bm{x},t)$ in a selected volume $V(\bm{x},t)$, centered around $\bm{x}$, at time $t$, and computing the mass-weighted attribute 
\begin{equation}\label{eq:observation}
    p(\bm{x},t) = \left< \frac{\sum_{n \in \mathcal{N}(\bm{x},t)} m_n p_n}{\sum_{n \in \mathcal{N}(\bm{x},t)} m_n} \right>_{t_\text{w}} ,
\end{equation}  
with a moving average $\left< \hdots \right>_{t_\text{w}}$ of time window $t_\text{w}$ for filtering the effects of coarse particles.

Each pseudo-particle has an individual size distribution, $\bm{f}_n = [f^1_n, f^2_n, \hdots, f^{N_\text{d}}_n ]$, discretized in $N_\text{d}$ size classes, where $f^k_n$ denote the mass percentage of particles in the size span $[d_{k-1},d_k]$.
The averaged size distribution in a volume $V(\bm{x},t)$ is computed using Eq.~(\ref{eq:observation}) to $\bm{f}(\bm{x},t)$ and the mean particle diameter by $\left<d\right>(\bm{x},t) = \bm{d} \cdot \bm{f}$, where $\bm{d}$ is the diameters of the of particle bins.
The cumulative size distribution $\bm{F}$ is the mass fraction of particles equal or larger than each size class $d_k$, such that $F^k = \sum_{k'=k}^{N_\text{s}} f^{k'}$.
The $D_F$ value is the diameter $d$ for which the cumulative size distribution equals $F$.
It is easily determined from the cumulative size distribution $F$ by direct inspection or interpolation of the nearest discrete size classes.

When particle simulation is needed for micro-tracking we use the discrete element method (DEM) with spherical pseudo-particles.
The simulation attributes include particle diameter $d_\text{p}$, coefficient of restitution $e_\text{p}$, friction coefficient $\mu_\text{p-t}$, rolling resistance coefficient $\mu_\text{p-r}$, and particle cohesion $c_\text{p}$.
The parameters are calibrated to values that approximate the mechanics and flow dynamics of the real material on bulk level, as measured by the bulk mass density $\rho_\text{b}$, internal friction $\mu_\text{b}$, cohesion $c_\text{b}$, and angle of repose.
The particle diameter is chosen as large as possible, but small enough for resolving the important flow features.
Even with pseudo-particles as coarse as a loader bucket, the number of particles in a silo or stockpile asset may exceed what can be simulated in realtime with present hardware.
To guarantee real-time performance of the particle dynamics, we employ data-driven model order reduction \cite{wallin:2021:ddm}.
The idea is to run numerous DEM simulations in advance, covering the relevant state-space as well as possible using authentic CAD models and control signals for in- and outflow to the asset.
From the simulation data, a low-dimensional model is trained to predict the flow field that occur in a stockpile or silo as a function of the control signal and mass distribution.
The computationally intense process of computing all contact forces is substituted by a simple evaluation of the predicted velocity field at the positions of the particles, which are simply advected using this velocity.
This is applied to the particles in the bulk. 
To accurately capture the dispersion of particles impacting the pile surfaces, resolved DEM is used. 
The combination of the two models in an executable application, part of a distributed system, is referred to as granular surrogates. 

\section{Integration test}
\begin{figure}[H]
    \centering
    \centering
  \captionsetup{width=0.85\textwidth}
  \includegraphics[width=0.85\textwidth]{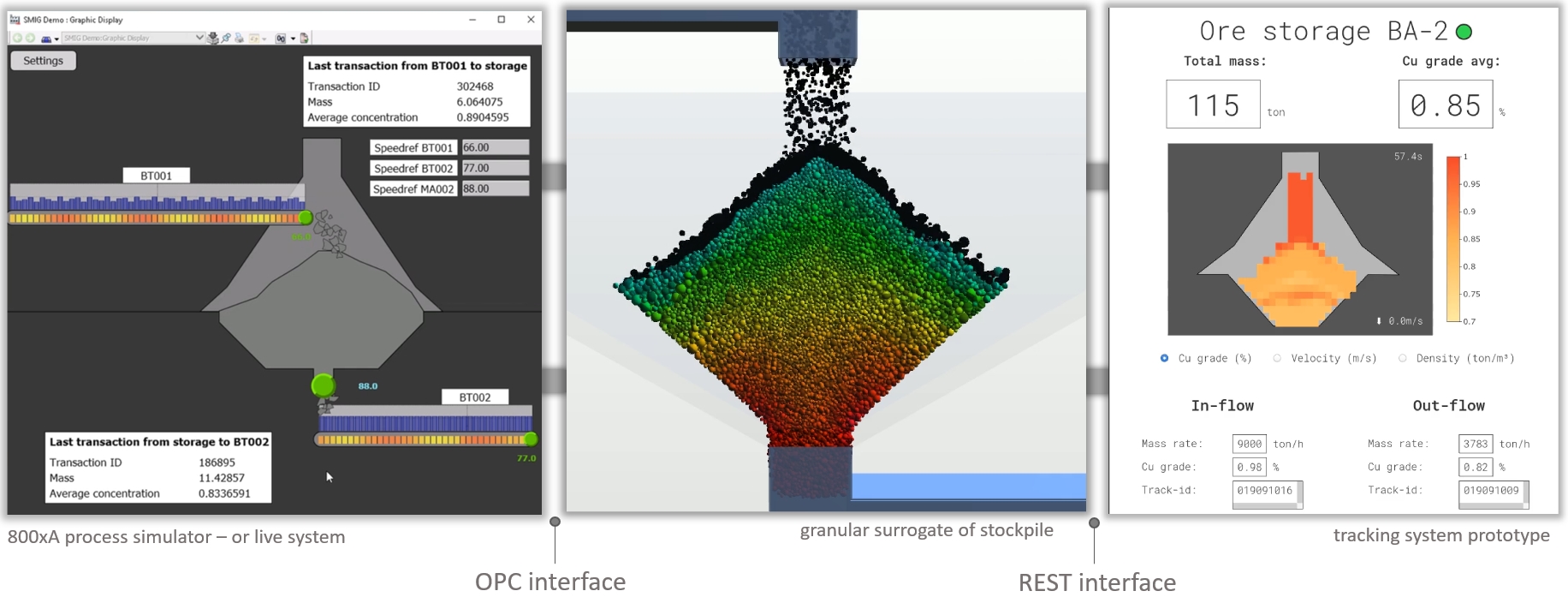}
    \caption{The integration test system include two conveyor assets in an ABB 800xA simulator (left), a storage asset (centre) driven by a granular surrogate, and a monitoring service (right) displaying macro-tracking data and virtual sensor data from the storage. 
    The communication interfaces rely on OPC and REST, respectively.
    In the granular surrogate, black particles are dynamic while the others are propagated using the reduced model and color coded by residence time. A video is included as supplementary material S1.}\label{fig:integration_test}
\end{figure}

An integration test of using granular surrogates for micro- and macro-tracking was performed, see Fig.~\ref{fig:integration_test}.
A virtual stockpile asset was created using AGX Dynamics for the particle simulation.
To accelerate the simulations to realtime performance, a reduced order model was trained from groundtruth simulation data, as described in \cite{wallin:2021:ddm}, with tracking accuracy of 10\%.
The storage asset was connected to an ABB 800xA process simulator \cite{ABB800xA} using OPC \cite{OPC-UA} for receiving a discharge control signal and for exchanging macro-tracking data.
The ABB simulator represent the assets of the incoming and outgoing conveyors, discretized in bins moving with the controlled conveyor belt speed. 
Each bin on the incoming conveyor contain material with a specific mass, ore grade and a tracking-id.
That information is exchanged with the stockpile asset when a bin reach the discharge point. 
The storage asset responds by emitting particles with the corresponding total mass and mineral concentration.
The particles fall, impact, and disperse over the pile surface as simulated by the physics-based stockpile model.
The storage asset receives a discharge rate control signal from the ABB simulator.
This drives a flow field in the bulk and particles are propagated in realtime by the data-driven reduced order model.
The particles that reach the outlet are eliminated and a macro-tracking message is passed so that the ABB simulator receive information of the material that enters the outgoing conveyer.
That information include tracking-ids, mass and ore grade.
The micro-tracking data resides in the storage asset. 
A monitoring service is created as a stand-alone web application. 
It receives virtual sensor information from the stockpile asset using a REST\footnote{Representational state transfer (REST): \url{https://en.wikipedia.org/wiki/Representational_state_transfer}} interface.
The monitoring information include the total mass in the storage, the average grade, and the spatial distributions of these quantities as well as the predicted velocity field in the bulk of the pile.
A video that demonstrate the integration test is included as supplementary material S2.
It should be pointed out that the ABB 800xA simulator can be switched to sharing realtime data from a mine with the same layout.

\section{Simulator}
\begin{figure}[H]
    \centering
    \captionsetup{width=0.85\textwidth}
    \includegraphics[width=9 cm]{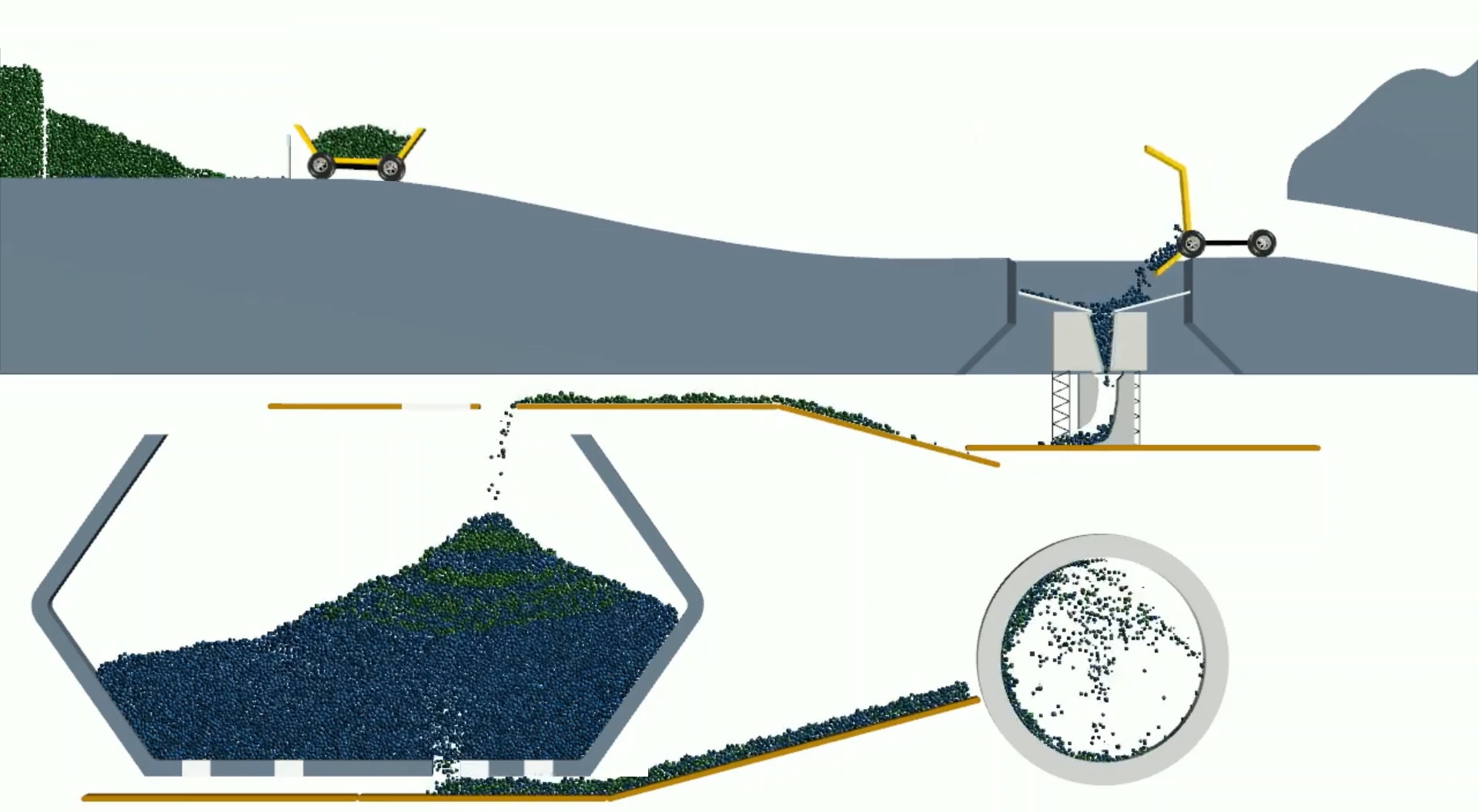}
    \caption{Image from the simulator with material hauled with trucks from two sources, dumped at a jaw crusher, conveyed to and from a storage, and finally fed to a grinder.
    Green particles come from the left source and blue particles from the right source.
    A video is available as supplementary material S2.
    \label{fig:simulator}}
\end{figure}

Before implementation in a real mine, particle-based material tracking is tested in a simulator of a simplified 2D model of a mine shown in Fig.~\ref{fig:simulator}. 
A video of the simulator is included as supplementary material S2.
Each asset can easily be replaced with more elaborate 3D models or virtual assets mirroring real physical assets. 
There are two sources of ore from which haul trucks transport blasted material to a single crusher. 
The crushed material is fed to a stockpile for intermediate storage, from which material is drawn and fed to a mill.
Belt conveyors transport the material between the crusher, storage, and the mill.
The simulator is used for testing the analysis that is enabled by the particle-based tracking and for understanding the sensitivity to tracking errors.

Macro-tracking events include the haul trucks dispatching from the loading point, dumping at the crusher station, and material passing the crusher zone, the inlets and outlets to the stockpile, and the inlet to the grinder.
Model order reduction is applied on the belt conveyors.
The simulator is implemented using the physics engine AGX Dynamics \cite{AGX21}, using a nonsmooth (or time-implicit) DEM \cite{servin:2014:esn} for the particles, and rigid multibody dynamics for the asset's moving parts using feedforward controllers. 
The pseudo-particle size range between $0.2$ and $0.48$ m, which corresponds to a mass range between $12$ to $160$ kg. 
The distance between the two loading points is roughly $150$ m and the stockpile is $60$ m tall.  
The simulations were run with time-step $11$ ms and $100$ projected Gauss-Seidel solver iterations, which corresponds to an error tolerance around $5$ \% for the stockpile bulk behavior \cite{servin:2014:esn}.

\begin{figure}[H]
    \centering
   \captionsetup{width=0.85\textwidth}
    \includegraphics[width=6 cm, trim = 0 0 0 0, clip]{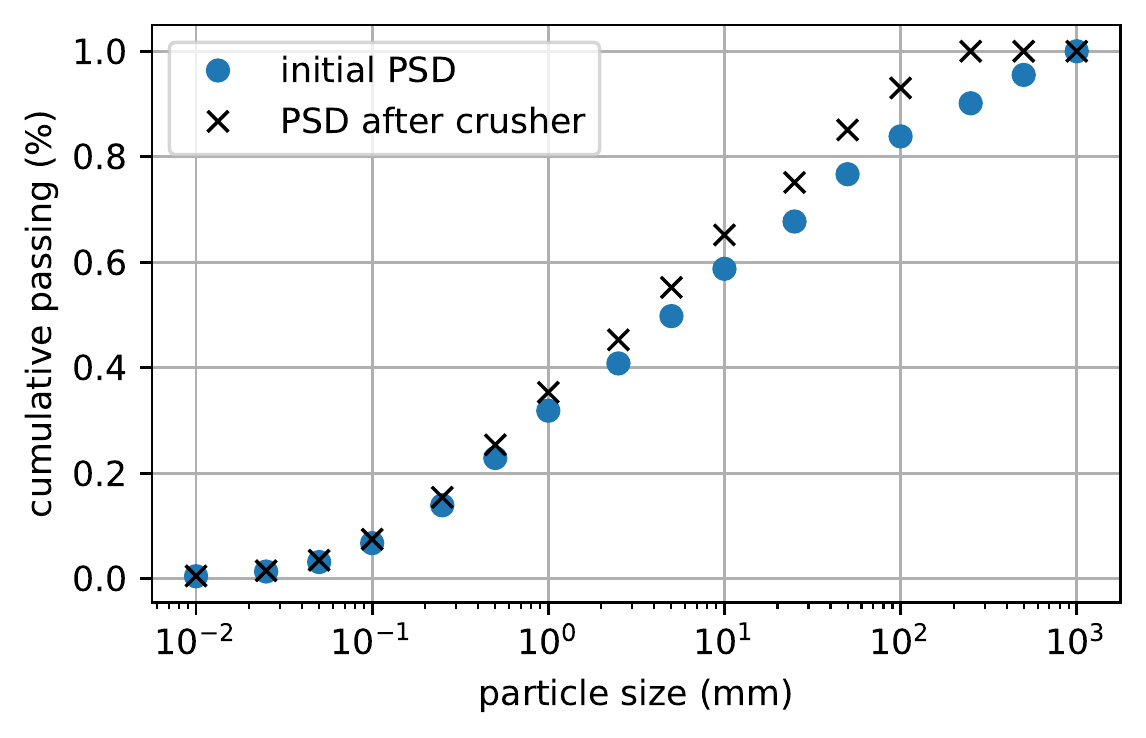}
    \caption{The cumulative particle size distribution (PSD) after blasting and after the crusher.}
    \label{fig:size_distribution}
\end{figure}

At the sources, each particle is assigned an identity and a location of origin.
The left source is assigned a mineral concentration of $0.2$, while the right source is assigned the value $0.8$. 
The right source is given a mineral hardness of $1.0$, while the hardness at the left source varies over the ore body according to some function $h(\bm{x})$. 
It is tested how a step in the hardness distribution function, from value $1.0$ to $2.0$, in the left source propagates through the system. 
It is also investigated how well a hardness profile in the ore body can be backtracked from observations at the crusher and the grinder. 
The (true) particle size distribution is discretized in $N_\text{d} = 16$ size classes with a logarithmic binning. The initial size distributions is shown in Fig.~\ref{fig:size_distribution}, including also the average size distribution after the crusher.
Do note that the pseudo-particle diameter, used for visualization and contact dynamics, have no correspondence to its internal particle size distribution.

Each haul truck receive a load of roughly $15$ ton (roughly $150$ pseudo-particles) at the loading point. 
For simplicity, the loading equipment is not modeled explicitly.
The trucks dump the load on the crusher inlet and there occur mixing of the material sources from here on.
The crusher is modeled as a jaw crusher, drawing material from above and crushing it with a jaw driven in an oscillatory motion. 
The pseudo-particles retain their original diameter, but when they pass the jaw, the internal particle size distribution is transformed.
The fractions larger than the crusher gap width, $d_\text{cr} = 0.25$ m, break into a uniform distribution of the smaller size classes. 
This unit operation is represented by $\bm{f}_n := \bm{f}_n - \Delta \bm{f}_{n-} + \Delta \bm{f}_{n+}$, with the change vector for crushed (oversize) fractions, $\Delta f^k_{n-} = f^k_n$ for $d_k \geq d_\text{cr}$ (otherwise zero),
and the change vector for the $n_\text{d}$ undersized fractions $\Delta f^k_{n+} = \tfrac{1}{n_\text{d}}\sum_{k'} \Delta f^{k'}_{n-}$ ( otherwise zero).
The resulting size distribution is shown in Fig.~\ref{fig:size_distribution}. 
We assume that the torque for driving the jaw crusher stands in proportion to the amount of mass in the crusher zone $V_\text{cr}(t)$, its hardness and size distribution, according to the following model
\begin{equation}\label{eq:crusher_torque}
    \tau_\text{cr}(t) = \sum_{n \in \mathcal{N}_{V_\text{cr}}} m_n h_n \bm{k}_\text{cr} \cdot \bm{f}_n  ,
\end{equation}
where the crushing coefficient $k^n_\text{cr} = 1.0$ for the oversize and $k^n_\text{cr} = 0.1$ for the undersize.
The momentaneous power draw of the crusher is computed by $P_\text{cr} (t) = \tau_\text{cr} \omega_\text{cr}$,
where $\omega_\text{cr} = 10$ rad/s is the crusher speed.
%

\begin{figure}[H]
    \centering
   \captionsetup{width=0.85\textwidth}
    \includegraphics[width=6 cm, trim = 0 0 0 0, clip]{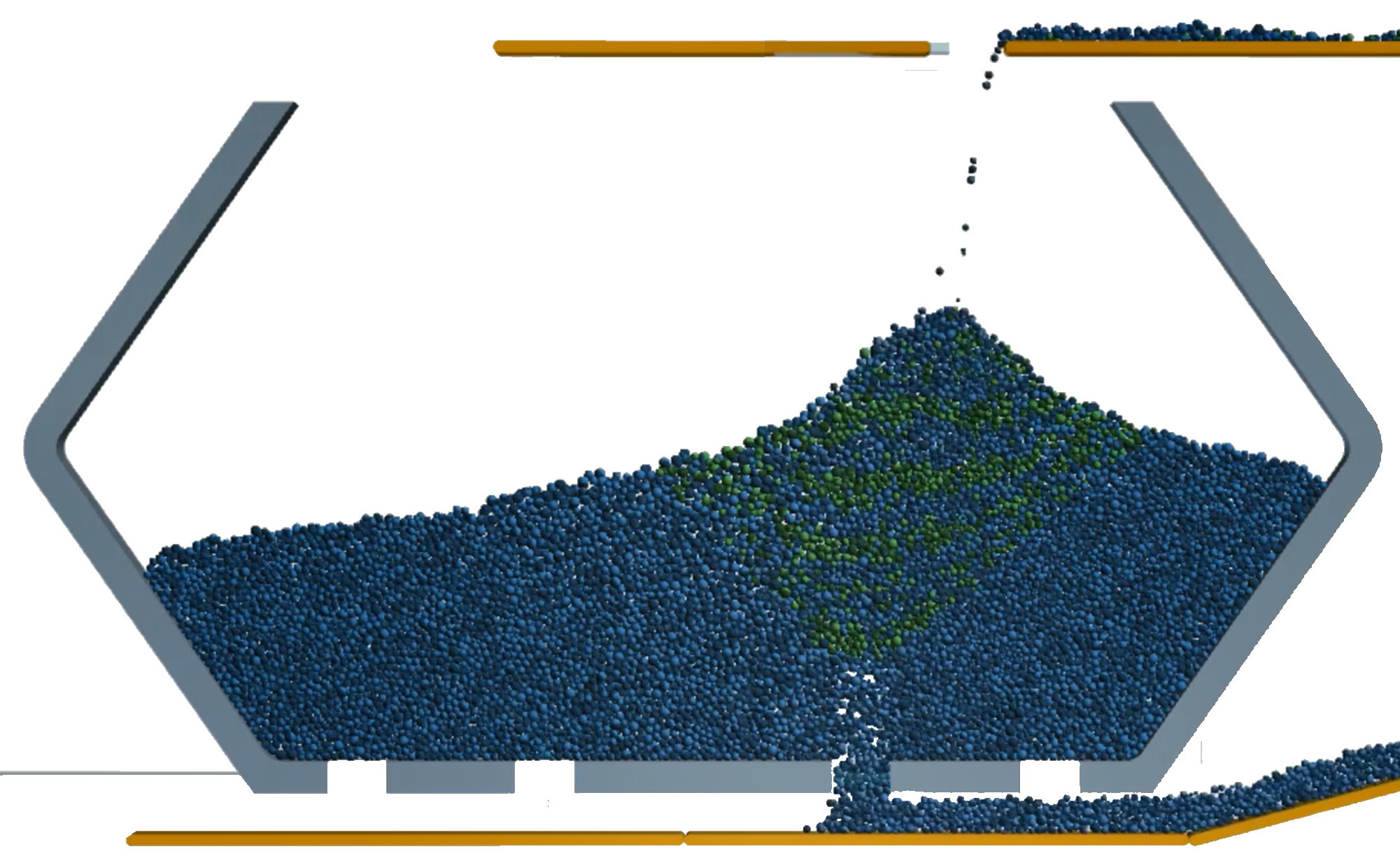}
    \hspace{5mm}
    \includegraphics[width=6 cm, trim = 0 0 0 0, clip]{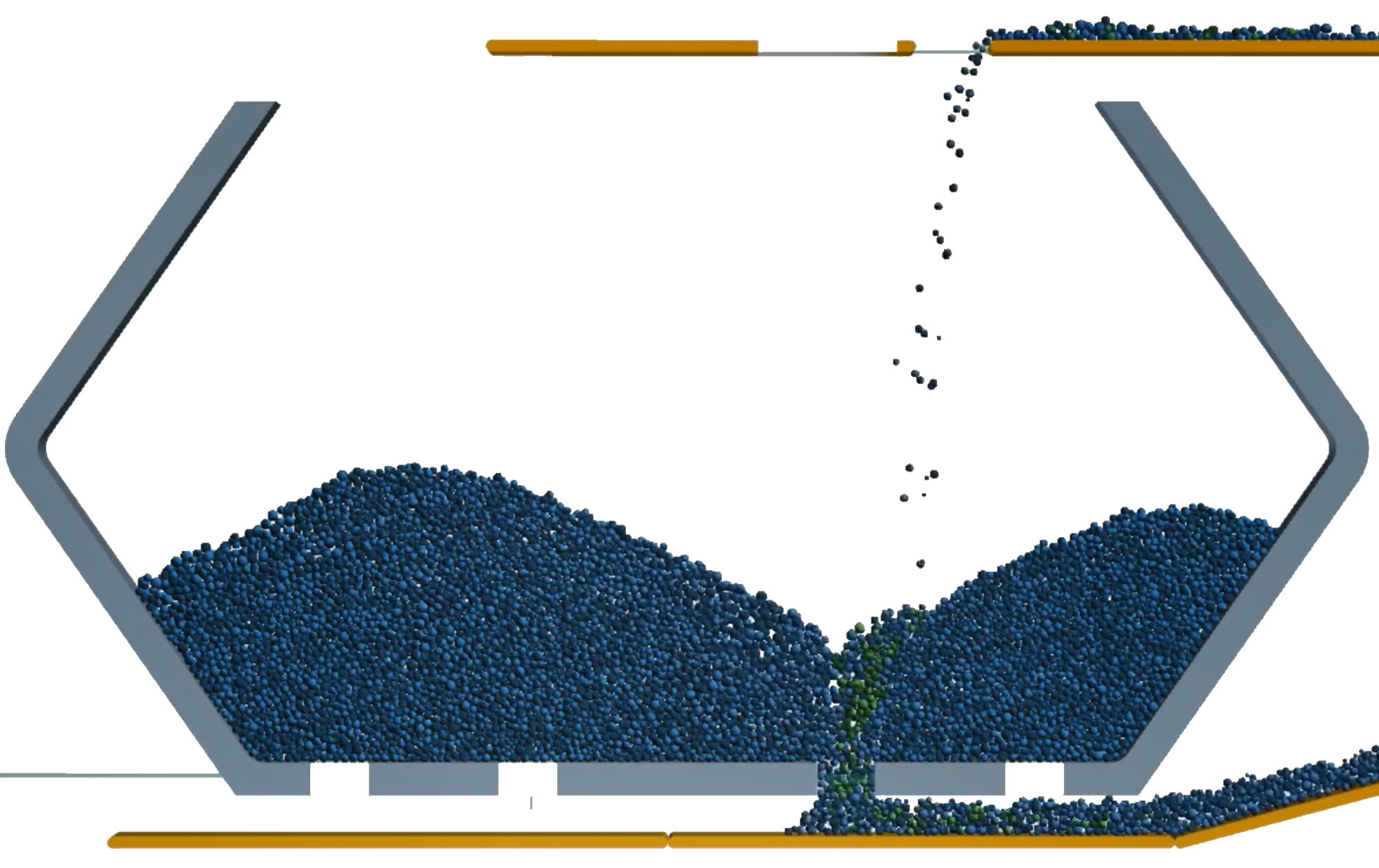}
    \caption{The storage asset at two different pile states, point (left) and depleeted (right).
    Blue and green particles have hardness value $1$ and $2$, respectively.}
    \label{fig:pile_states}
\end{figure}

The stockpile holds up to $2,000$ tons (roughly $20,000$ k pseudo-particles). 
There are two entry discharge points and four exit discharge points, but only one is used for the presented tests.
When material is discharged from the bottom outlet, at a controlled speed ranging up to $2$ m/s (or $2$ ton/s), a funnel flow field is activated in the pile, reaching the surface where a depletion is gradually formed, unless new material is fed into the storage at the same rate or faster.
Two time instances of pile states are shown in Fig.~\ref{fig:pile_states}.
For the pointy pile state one can expect more mixing of the incoming material and longer residence time, contrary to the depleted pile state where material flow more directly through the storage and with less mixing.

\begin{figure}[H]
    \centering
  \captionsetup{width=0.85\textwidth}
    \includegraphics[width=8 cm, trim = 0 0 0 0, clip]{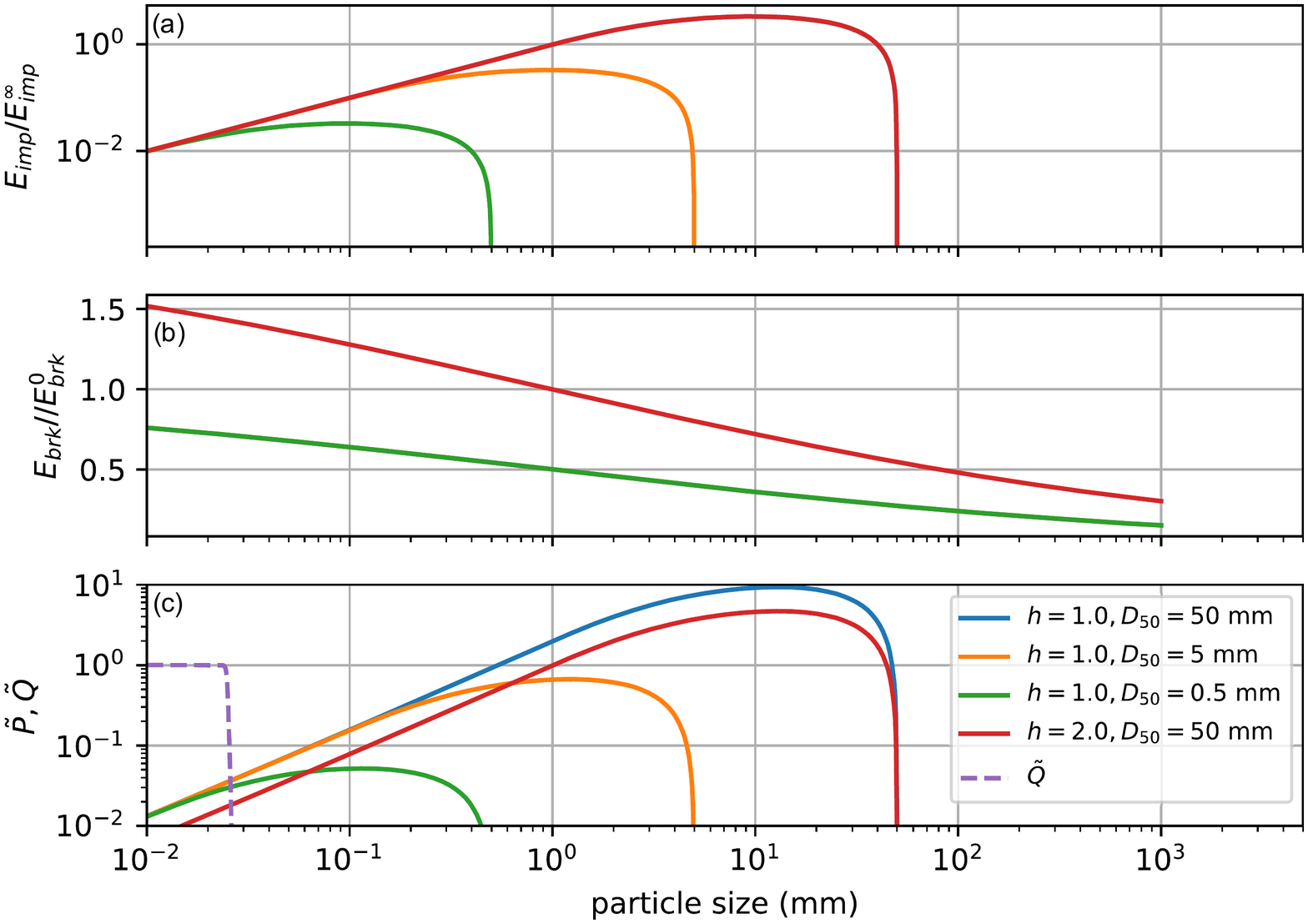}
    \caption{The mill breakage model as function of particle size and mineral hardness
    illustrated in terms of the impact energy distribution (a), specific breakage energy (b), and the breakage probability rate (c). The probability rate for exiting the mill is also included. }
    \label{fig:mill_breakage_model}
\end{figure}

An idealized mill model is assumed, guided by a model that has been used for mill control at Boliden Aitik \cite{Araker2020}. 
The idealized model consists of a single grinder, although most real mills consists of a circuit of grinders and classifiers.
When a pseudo-particle $n$ reach the the grinder, a size reduction model begins to operate on its size distribution. 
Each size fraction $k$ is assigned a probability per unit time, $\mathcal{P}^k_n$, for breaking into smaller fractions and a probability per unit time, $\mathcal{Q}^k_n$, for leaving the mill.
At each time-step the mass fractions carried by each pseudo-particle are redistributed according to $\bm{f}_n := \bm{f}_n - \Delta \bm{f}_{n-} + \Delta \bm{f}_{n+}$, where the broken and passing fractions are
\begin{equation}
    \Delta f^k_{n-} = ( \mathcal{P}^k_n + \mathcal{Q}^k_n) f^k_n \Delta t .
\end{equation}
and the reduced fractions are
\begin{equation}
    \Delta f^k_{n+} = \tfrac{1}{N_\text{d} - k} \sum^{N_\text{d}}_{k' = k + 1} \mathcal{P}^{k'}_n f^k_n \Delta t .
\end{equation}
Modeling a size-classifier, the probability per unit time for the finest fractions, smaller than $d_\text{cl}$, to leave the mill is $\mathcal{Q}^k_n = k_\text{esc} / (1 + \exp[k_\text{cl}(d_k/d_\text{cl} - 1)])$. We use $d_\text{cl} = 0.025$ mm, $k_\text{cl} = 10$, and $k_\text{esc} = 10^3$.
The breakage probability depends on the (true) particle size, impact energy, hardness, and the number of impacts \cite{Bruchmueller2011}, which in turn depends on the mill's angular velocity, $\omega_\text{gr}$, and the active torque $\tau_\text{gr-a}$.
We assume the following model for breakage probability per unit time
\begin{equation}
    \mathcal{P}^k_n = k_\text{gr} \frac{E^k_\text{imp}}{E^k_{n,\text{brk}}} 
    \left(\frac{\tau_\text{gr-a} \omega_\text{gr}}{\tau^0_\text{gr-dr}\omega_\text{crit}}
    \right)^2 ,
\end{equation}
where $E^k_\text{imp} = E^\infty_\text{imp} \left( e - \exp [1 - D_{50}/d_k] \right) d_k / d_\text{gr}$ is an impact energy distribution function, $E^k_{n,\text{brk}}  = E^0_\text{brk} h_n / [1 + (d_k / d_\text{gr})^{1/4}]$ is a specific breakage energy distribution, $D_{50}$ is the (time-dependent) mean diameter of the material currently in the grinder, $d_\text{gr} = 1.0$ mm is a grinding length-scale parameter, and $k_\text{gr} = 1.0$ a dimensionless breakage coefficient.
The internal drive friction, $\tau^0_\text{gr-dr}$, of an empty mill and the critical mill speed, $\omega_\text{crit} = 1.5$ rad/s, are used for normalization. 
The mill breakage model can be seen in Fig.~\ref{fig:mill_breakage_model}, with $\tilde{\mathcal{P}}(d_k) \equiv \mathcal{P}(d_k) /\tfrac{\tau_\text{gr-a} \omega_\text{gr}}{\tau_\text{mill}\omega_\text{crit}}$. 
The exit probability $\tilde{\mathcal{Q}}(d_k) \equiv \mathcal{Q}(d_k) / k_\text{esc}$ is also included.
The total grinding torque is the sum of three contributions, $\tau_\text{gr} = \tau_\text{gr-dr} + \tau_\text{acc} + \tau_\text{gr-a}$. The total mass, $m_\text{tot}(t) = m_0 + m_\text{gr}(t)$, is the mass of the moving part of the mill, $m_0 = 10^3$ kg, plus the mass of the material in the grinder, $m_\text{gr}(t) = \sum_n m_n$.
The internal drive friction is modelled $\tau_\text{gr-dr} = \mu_\text{dr} m_\text{tot} g r_\text{mill}$ and the torque required for accelerating the grinder is $\tau_\text{acc} = m_\text{tot} r^2_\text{mill} \dot{\omega}_\text{gr}$, where we denote the internal friction coefficient $\mu_\text{dr} = 0.02$, gravity acceleration $g = 9.8$ m/s$^2$, and mill radius $r_\text{mill} = 5$ m.
The active torque depends on the charge's mass, $m_\text{gr} = \sum_n m_n$, size distribution, and dynamic angle, which depends on the mill angular velocity.
We assume, again guided by \cite{Araker2020}, that the active torque can be modeled $\tau_\text{gr-a} = \mu_\text{gr-a} m_\text{gr} g r_\text{mill}$ with effective internal friction of the charge
$\mu_\text{gr-a} = k_\text{gr-a} \Theta(\omega_\text{gr}/\omega_\text{crit}) \tfrac{m_\text{gr-a}\omega_\text{gr}}{m_\text{gr}\omega_\text{crit}}$.
Here, $m_\text{gr-a} = \sum_{n,k\geq N_\text{gr-a}} m_n^k$ is the mass of the charge fractions larger than an active grinding size $d_\text{gr-a}$, with index $N_\text{gr-a}$.  
We use $d_\text{gr-a} = 1.0$ mm, which is $N_\text{gr-a} = 7$ in the 16 class size representation, and $k_\text{gr-a} = 1$. 
With $\Theta(x) = \operatorname{Re}[(1 - x^4)^{1/2}]$ the effective internal charge friction and active torque drops to zero at and above the critical velocity.
The power draw of the grinder is computed $P_\text{gr} (t) = \tau_\text{gr} \omega_\text{gr}$.
When nothing else is stated, the mill is run at constant nominal speed $\omega_\text{gr} = 1.2$ rad/s.
\section{Results}

\subsection{Step response}
\begin{figure}[H]
    \centering
  \captionsetup{width=0.85\textwidth}
    \includegraphics[width=8 cm, trim = 0 0 0 0, clip]{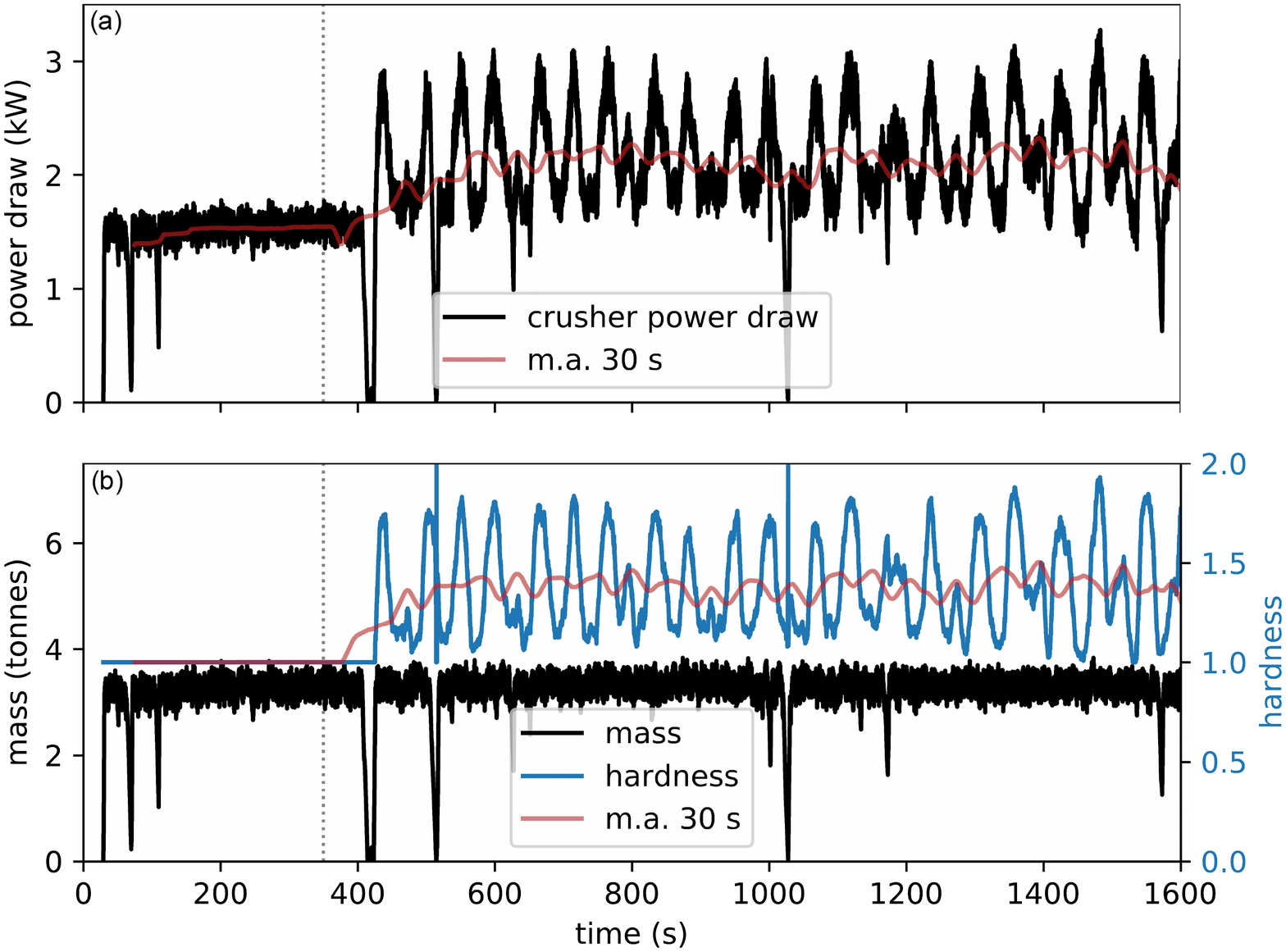}
    \caption{The response in the crusher power draw (a) from a step change in the mineral hardness (b). Included are also $30$ s moving averages and the momentaneous mass in the crusher volume is also shown.}
    \label{fig:crusherStep}
\end{figure}

\begin{figure}[H]
    \centering
  \captionsetup{width=0.85\textwidth}
    \includegraphics[width=8 cm, trim = 0 0 0 0, clip]{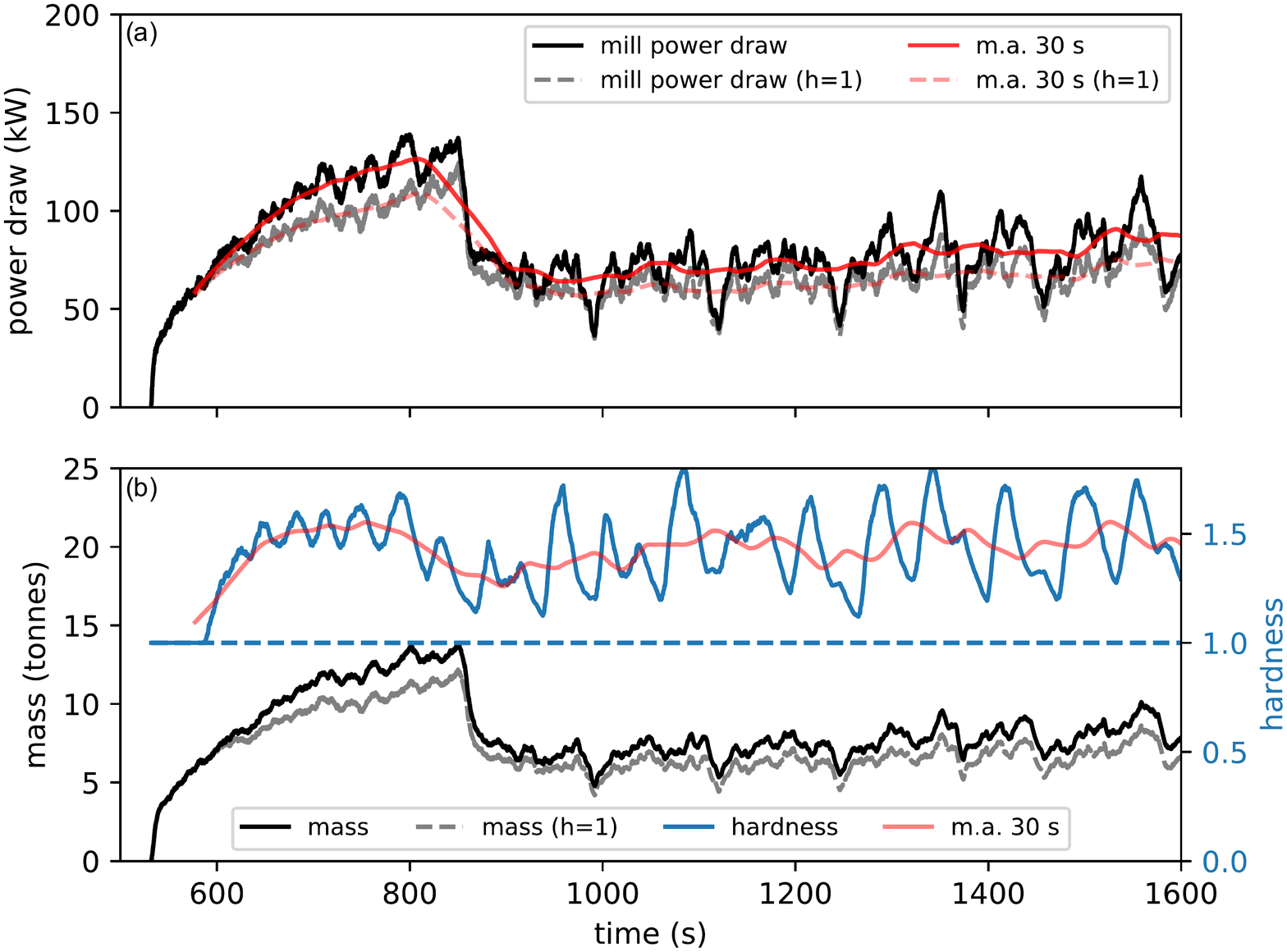}
    \caption{The response in the mill power draw (a) from a step change in the ore hardness (b) after passing through the storage in Fig.~\ref{fig:pile_states}. 
    The ore mass in the mill is also included and a $30$ s moving average of the power draw and hardness.
    .  For reference, the case of constant hardness ($h=1$) is also included.}
    \label{fig:millStepBest}
\end{figure}

The simulator is initialized with ore of constant hardness $h = 1$ transported from both sources.
A step change to hardness $h = 2$ is introduced at the left source and the response in the
crusher may be seen in Fig.~\ref{fig:crusherStep}.
The haul truck reach this harder material at time $t = 350$ s. 
About $60$ s later (five seconds more than one truck cycle) the crusher power draw has increased by $40$\%, from $1.5$ kW to $2.1$ kW.
A virtual sensor in the crusher reveal that the mineral hardness varies between 1 and 2, with a time-averaged hardness around $1.4$.
The large oscillations in hardness and power draw, with $55$ s time period, is because of the trucks dumping intechangely from the left and right source with hardness $1$ and $2$, respectively.
For the step response in the mill we analyse the effect of the ore passing through the stockpile in Fig.~\ref{fig:pile_states}.
The discharge of the piled material start at $500$ and the feed reach the empty mill shortly thereafter.
The material in the mill remain of hardness $h=1$ for $90$ more seconds until it starts increasing at time $590$ s up to $h = 1.5$ with only small oscillations (well mixed). 
At time $875$ s, the storage level is down at zero (apart from the dead zones) and the rate of mass flowing into the mill drops quickly to the rate of mass leaving the crusher and flowing into the storage.
Therefore, after $875$ s the oscillations in hardness and power are again reflecting the truck cycles.
For reference we run the identical simulation with constant hardness $h=1$ at both sources.
As seen in Fig.~\ref{fig:millStepBest}, the difference in the two power draw signals follow the increase in material hardness inside the mill. The increase is about 20 \% on average.
Note that the mechanism for the higher power draw is that the harder ore takes longer time to grind and the mill therefore accumulate more mass over time.
%

\subsection{Identification of hardness from measurement-while-drilling}
\begin{figure}[H]
    \centering
  \captionsetup{width=0.85\textwidth}
    \includegraphics[width=6 cm, trim = 0 0 0 0, clip]{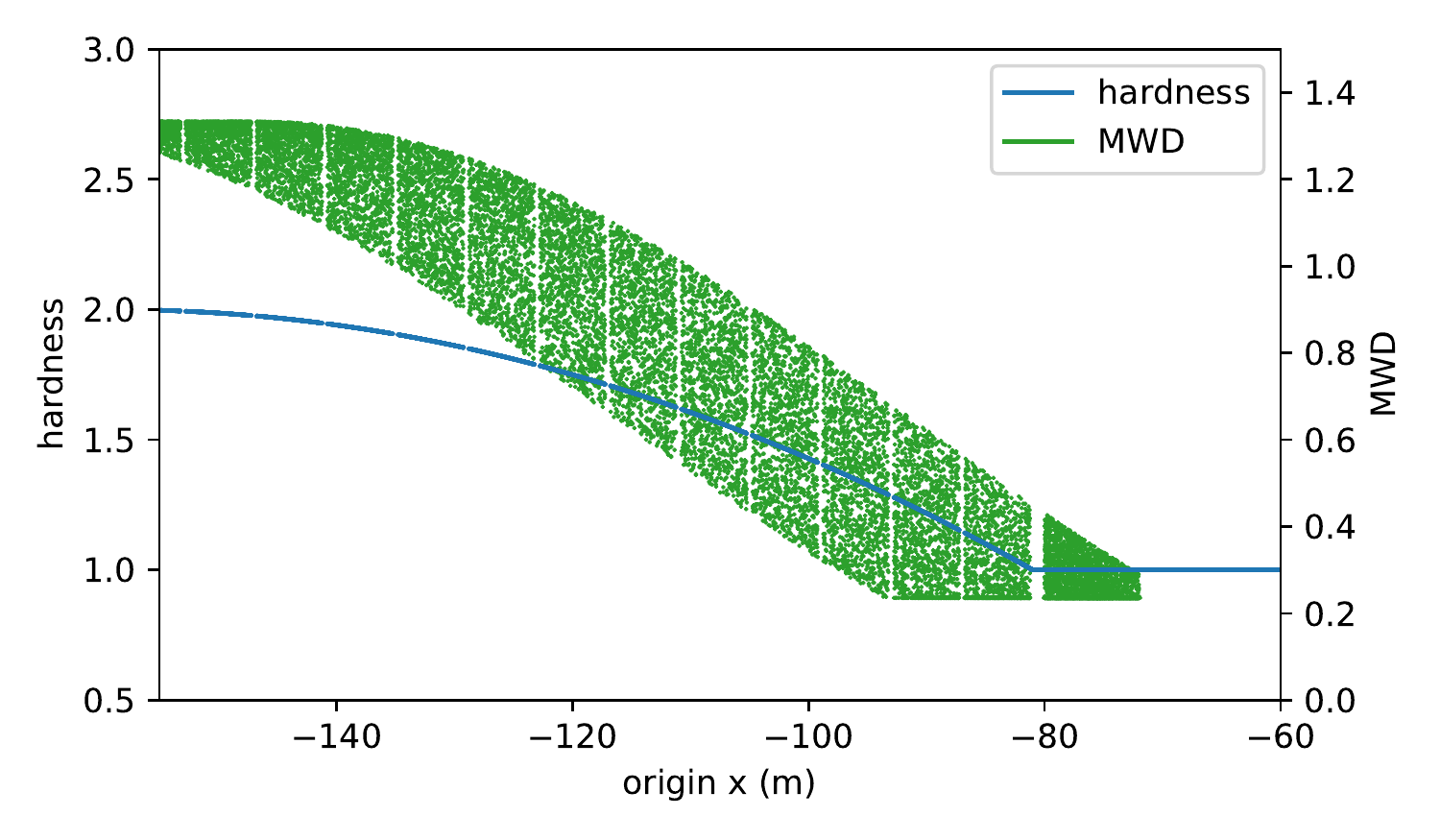}
    \caption{The assumed hardness profile (blue) at the left source and the corresponding MWD signal (green) after introducing a 12.5 m measurement location error at the source.}
    \label{fig:hardness_vs_MWD_tracking_error}
\end{figure}

\begin{figure}[H]
    \centering
  \captionsetup{width=0.85\textwidth}
    \includegraphics[width=12 cm, trim = 0mm 0mm 0mm 0mm, clip]{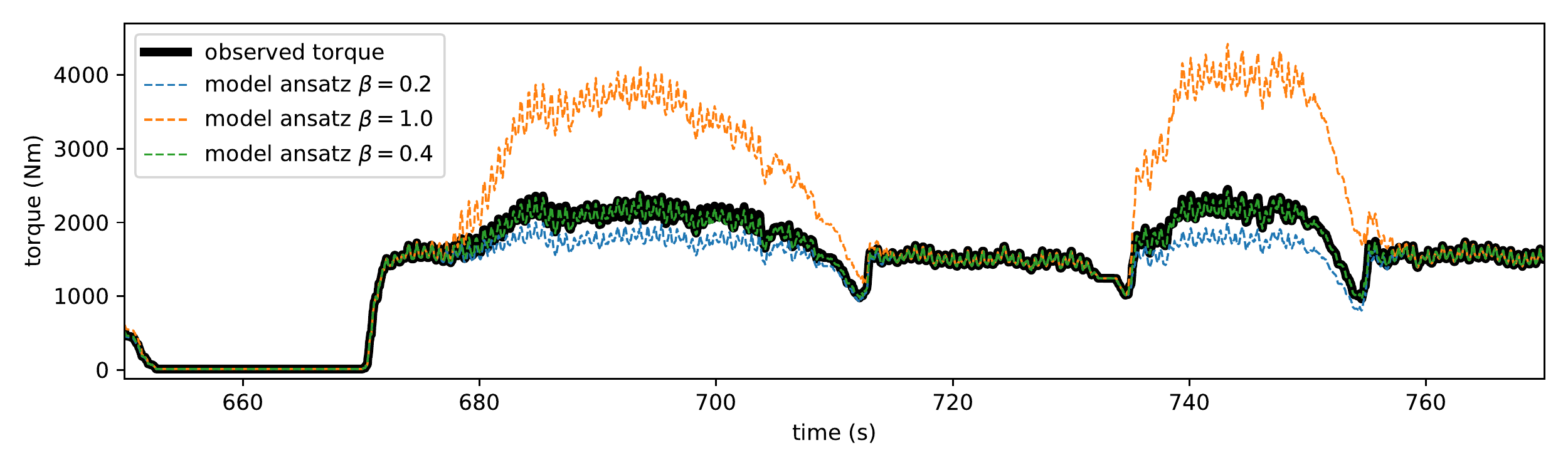}
    \caption{The observed crusher torque compared to the expected torque from different model ansatz given the MWD signal and hardness. 
    Four truck load reach the crusher during the shown time period, two from the left source with harder ore and two from the right source.}
    \label{fig:crusher_observed_torque_models}
\end{figure}

\begin{figure}[H]
    \centering
  \captionsetup{width=0.85\textwidth}
    \includegraphics[width=12 cm, clip]{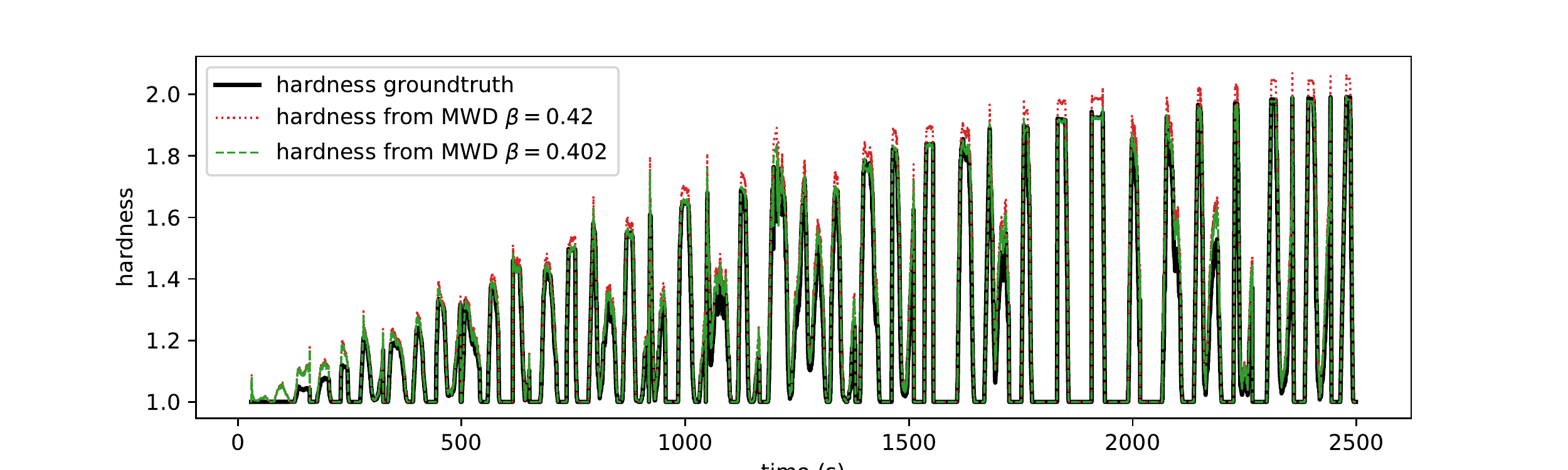}
    \caption{The true and predicted hardness given the MWD signal with the presence of a $12.5$ m tracking error. In addition to the best fitted model ($\beta = 0.402$) a slightly erroneous fitted model ($\beta = 0.42$) is included.}
    \label{fig:crusher_torque_to_hardness}
\end{figure}

To test the feasibility of identifying the rock hardness from the MWD signal we assume the quadratic hardness profile $h(x)$ in Fig.~\ref{fig:hardness_vs_MWD_tracking_error} at the left source, ranging continuously between $1$ and $2$.
While drilling, a measurement profile $s(x)$ is measured.
For the test we assume $s = h^{0.4}$ and introduce a measurement location error such that $s_n = h(x_n + \mathcal{R}(-\Delta x,\Delta x))^{0.4}$, where $\mathcal{R}(-\Delta x,\Delta x)$ is a uniform disturbance. We use $\Delta x = 12.5$ m, which corresponds to $6$ haul truck cycles back and fort to the crusher.
The right source has constant hardness $h = 1$.
Ore from the two sources enter the crusher, partially mixed, and give rise to the torque signal in Fig.~\ref{fig:crusher_observed_torque_models}.
Assuming the crusher torque is related to rock hardness according to Eq.~(\ref{eq:crusher_torque}) any ansatz $s(h,\beta)$ can be explored and optimized with respect to model parameter $\beta$.
Different outcomes for the ansatz $s = h^\beta$ are shown in Fig.~\ref{fig:crusher_observed_torque_models}. 
The best fit was found at $\beta = 0.402$, which deviate by $0.5$ \% from the ground truth because of the tracking location error that was introduced. 
Once a relation between the MWD signal and hardness has been established it is possible to predict the hardness of the material that is about to enter the crusher given the MWD signal measured at its location of origin.
This is illustrated in Fig.~\ref{fig:crusher_torque_to_hardness} over a sampling duration of $2,500$ s, or roughly 40 haul trucks from the left source.

\subsection{Predicting the properties of the mill feed}
\begin{figure}[h]
    \centering
  \captionsetup{width=0.85\textwidth}
    \includegraphics[width=0.225\textwidth, trim = 0 0 0 0, clip]{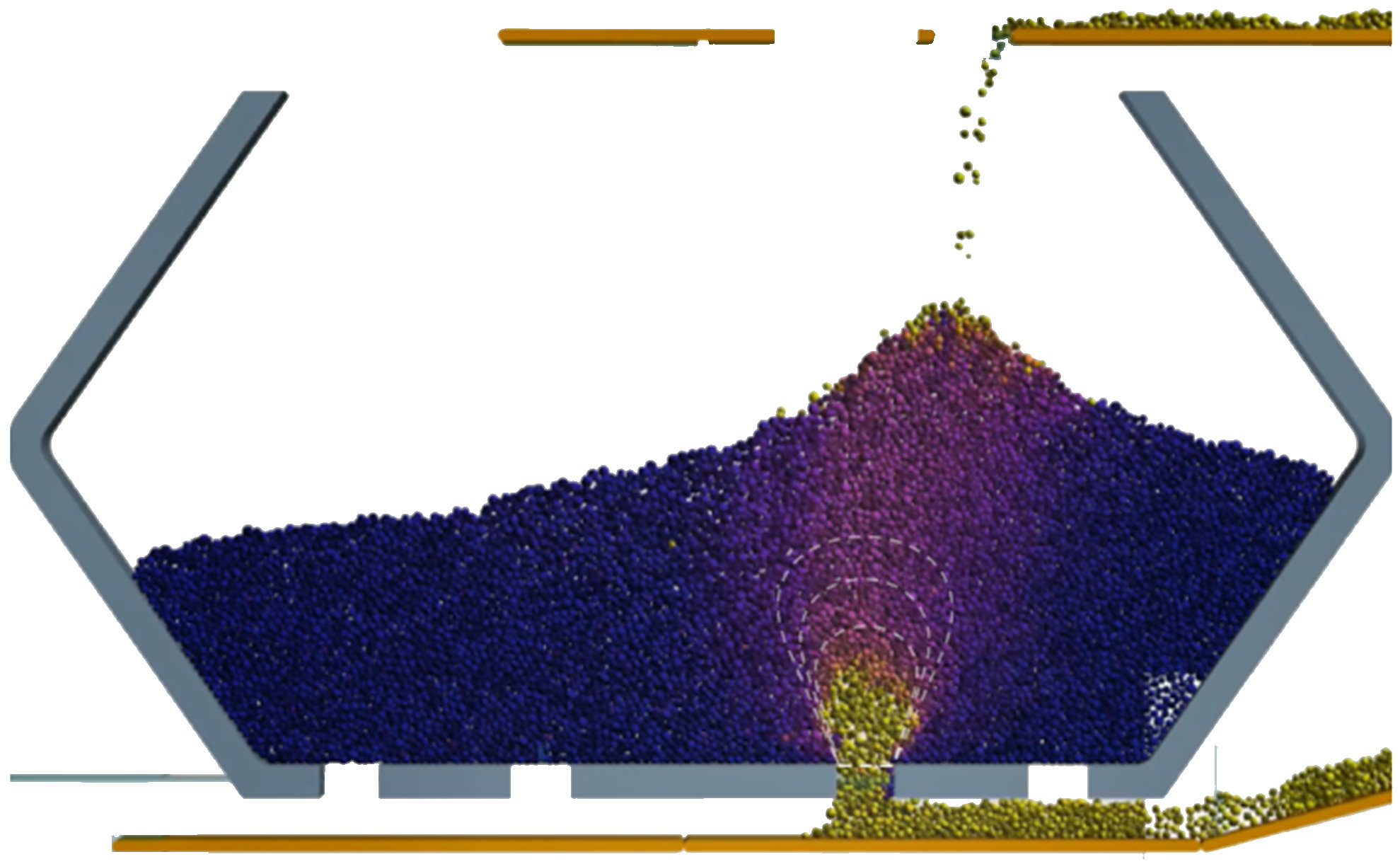}
    \includegraphics[width=0.225\textwidth, trim = 0 0 0 0, clip]{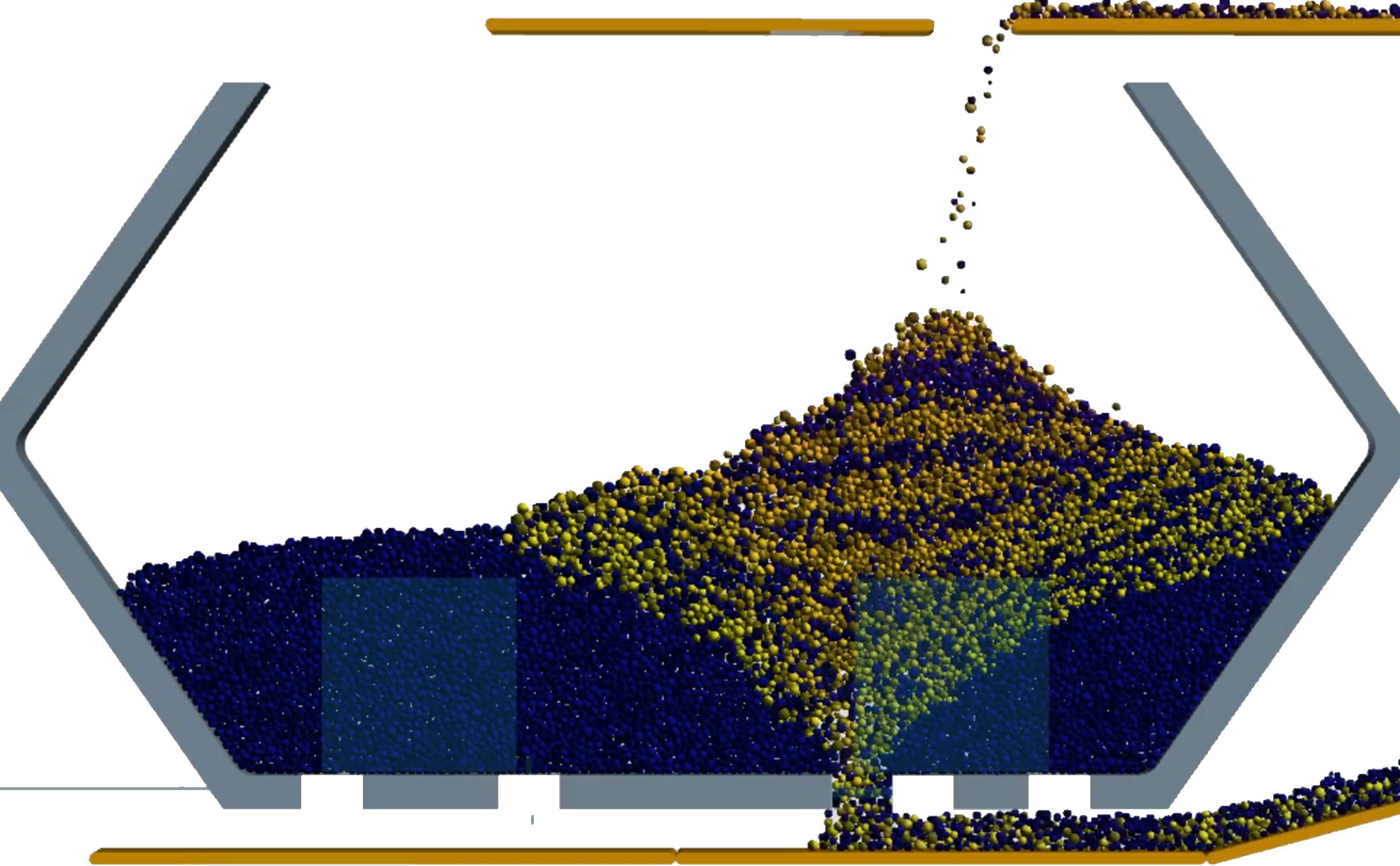}
    \includegraphics[width=0.225\textwidth, trim = 0 0 0 0, clip]{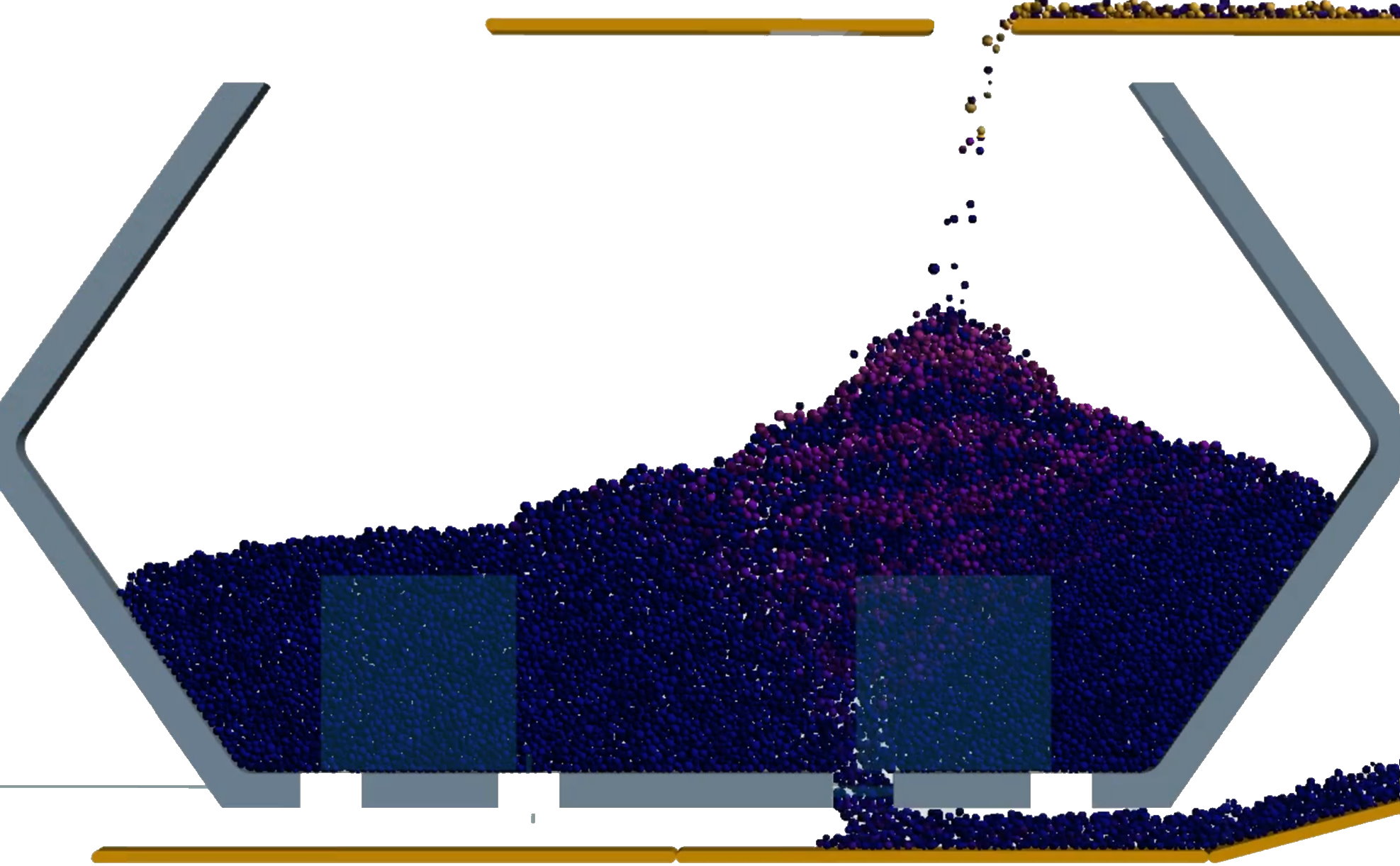}
    \caption{The spatial distribution of velocity (left), concentration (centre) and hardness (right) in the pile at a time instant.  The material that will be discharged in near time may be estimated from the funnel flow zone (left).}
    \label{fig:pileDistribution}
\end{figure}

\begin{figure}[h]
    \centering
  \captionsetup{width=0.85\textwidth}
    \includegraphics[width=0.35\textwidth, trim = 0 0 0 0, clip]{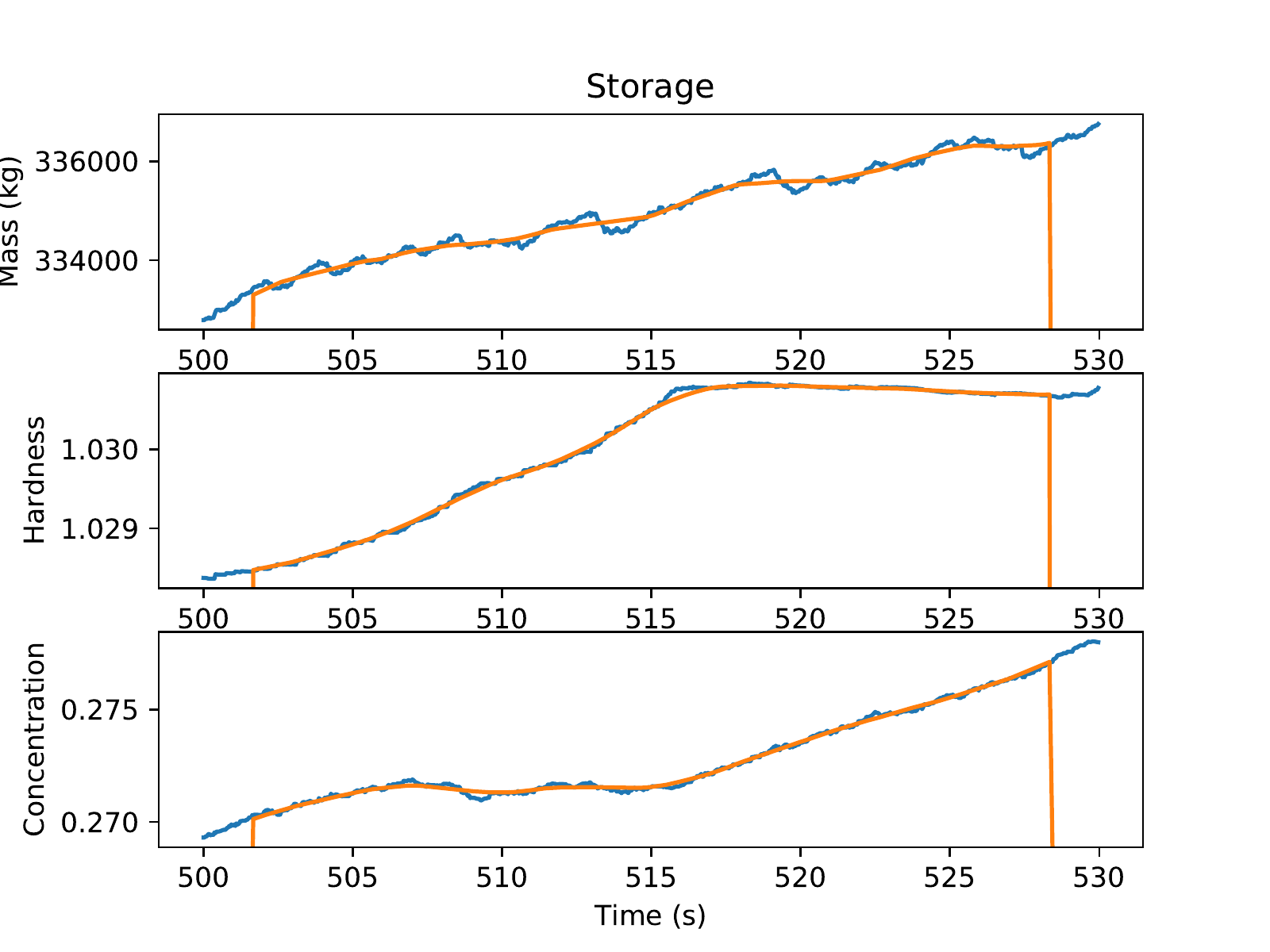}
    \includegraphics[width=0.35\textwidth, trim = 0 0 0 0, clip]{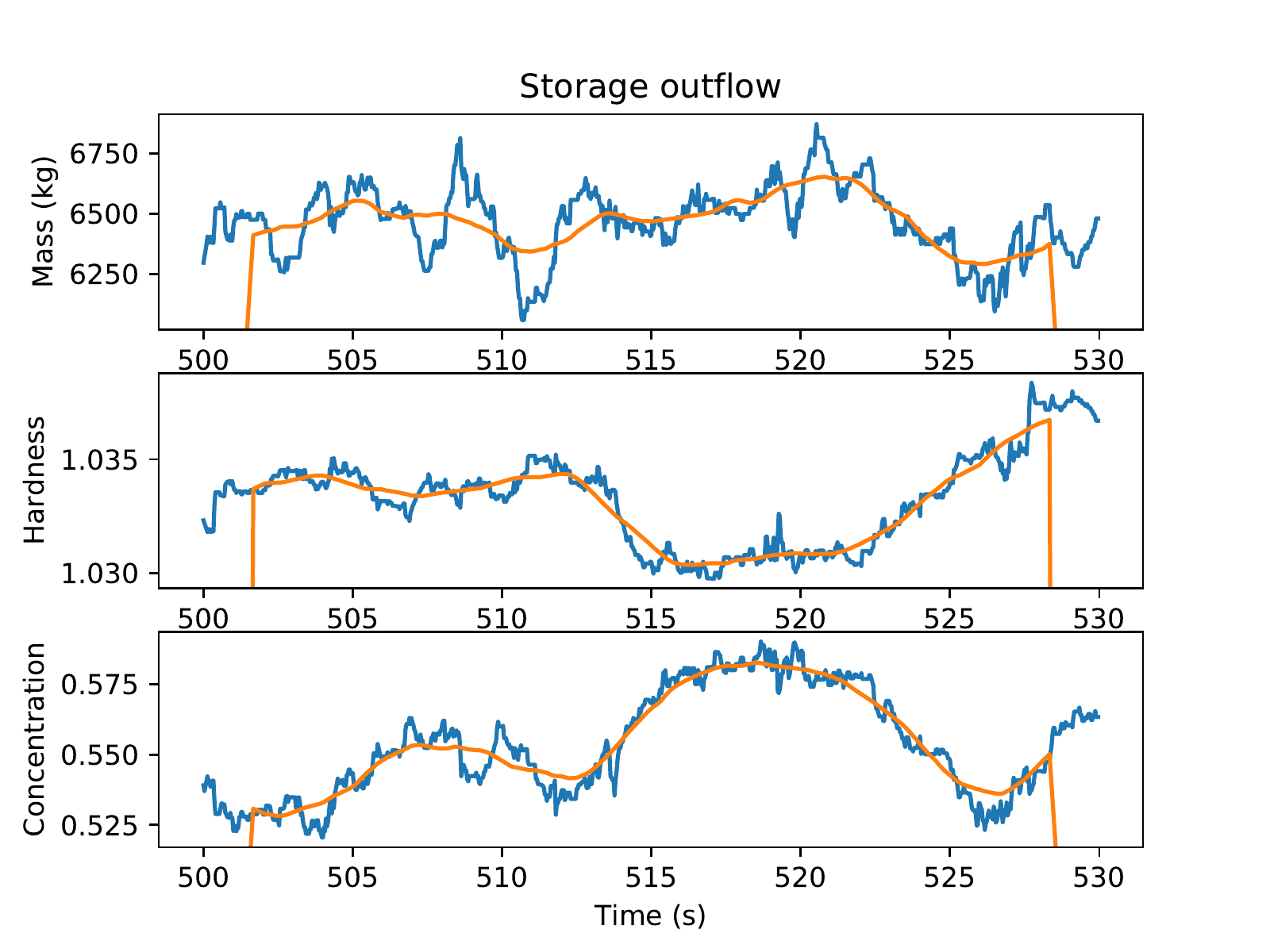}
    \caption{Evolution of the storage average properties over time (left) and the predicted feed for the next 30 s (right) given the funnel flow zone in Fig.~\ref{fig:pileDistribution}. A $2$ s moving average is included (orange).}
    \label{fig:storageEvolutionandOutflow}
\end{figure}

With particle-based tracking and Eq.~(\ref{eq:observation}) is straight-forward to monitor the  properties of the ore kept in the storage as function of time and space.  
Given a storage flow model it is also possible to estimate which particles in the funnel zone that will leave the storage in the near future, say the next $30$ s. 
This way the properties of the ore feed that will reach the mill can be predicted in advance, even before exiting the storage. 
Sample spatial distributions are shown in Fig.~\ref{fig:pileDistribution}. 
The time evolution of the properties in the pile and of the future feed are shown in Fig.~\ref{fig:storageEvolutionandOutflow}.

\subsubsection{Backtracking the grindability}
\begin{figure}[h]
     \centering
  \captionsetup{width=0.85\textwidth}
     \includegraphics[width=0.5\textwidth, trim = 0 0 0 0, clip]{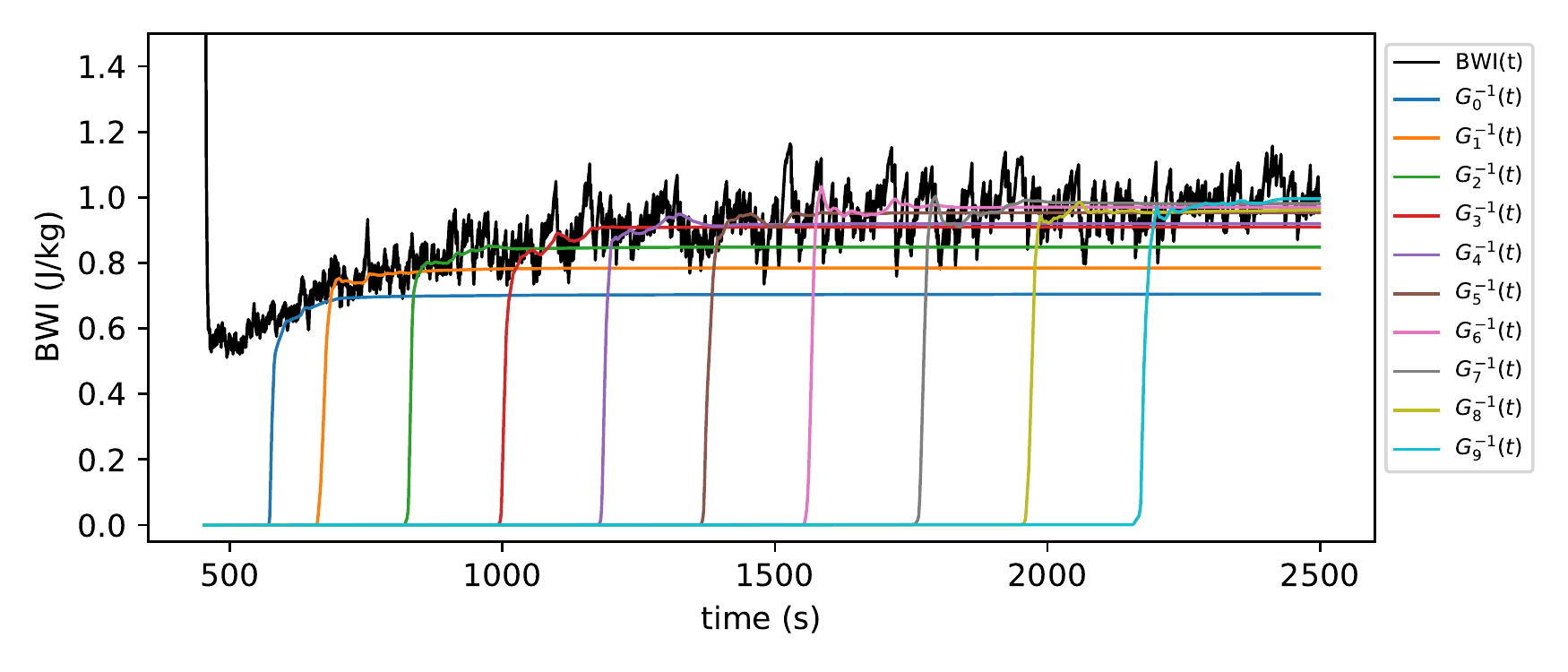}
     \caption{Time evolution of the Bond ball mill work index (BWI) measured in the mill with material from both sources.  The evolution of the backtracked (inverse) grindability at 10 individual blocks at the left source is also shown.
     The variations in power draw and hardness with $55$ s time period match the frequency of the truck hauling from the left source. 
     }
     \label{fig:grindability}
\end{figure}

\begin{figure}[h]
    \centering
  \captionsetup{width=0.85\textwidth}
    \includegraphics[width=0.5\textwidth, trim = 0 0 0 0, clip]{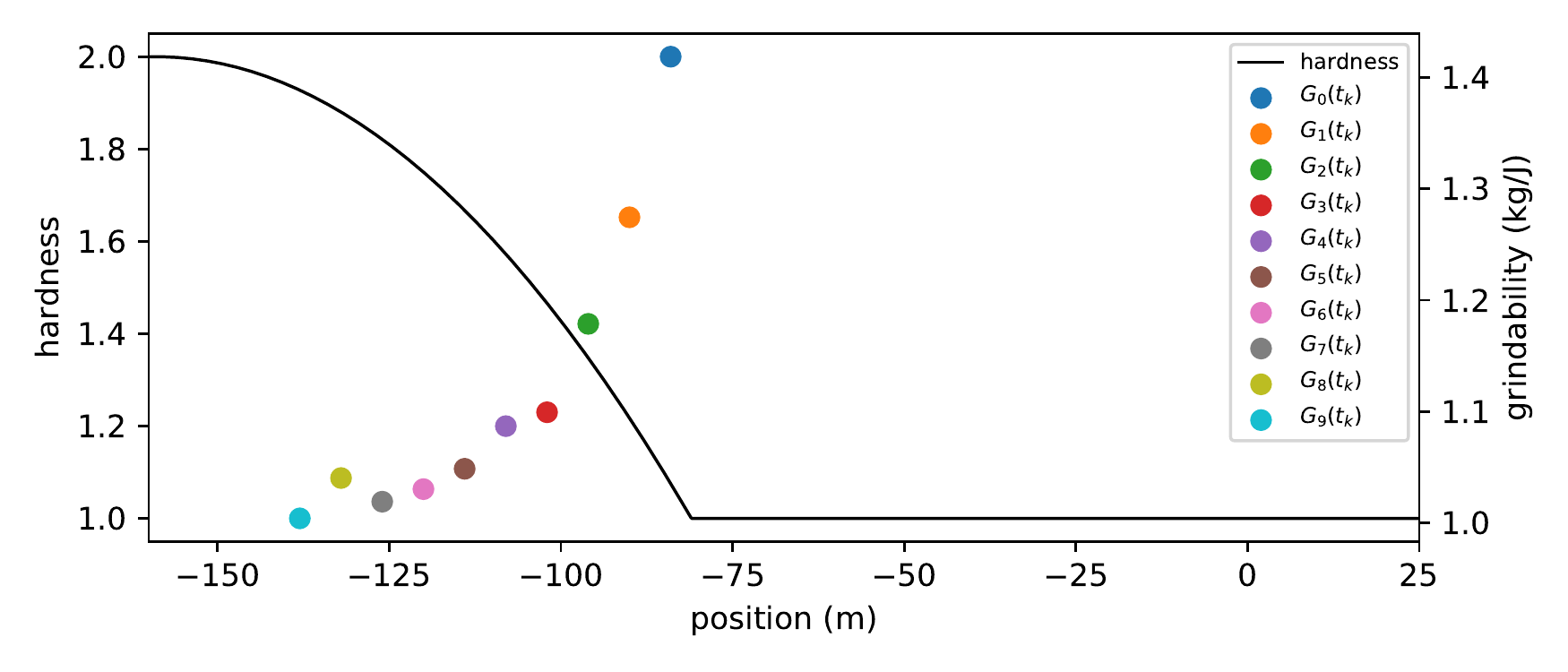}
    \caption{The backtracked grindability at 10 individual blocks at the left source shown together with the hardness profile. As can expected these are inversely related.}
    \label{fig:block_grindability}
\end{figure}

The grindability of an ore may be measured by the amount of energy, per unit mass, that is required for reducing the fragmentation down to a specific particle size.
For the sake of production planning the grindability is often estimated in advance using the relatively sparse exploration data of the ore body. 
The \emph{Bond ball mill work index} (BWI) is defined
\begin{equation}
    W = \frac{\text{BWI}}{3.6} \left[ \frac{10}{\sqrt{\left< D_{80}^{\text{out}}\right>}} - \frac{10}{\sqrt{\left< D_{80}^{\text{in}}\right>}}\right] ,
\end{equation}
where the work per unit mass (W/kg) is computed
$W(t) = \left< P_\text{gr} \right> / \left< \dot{m}_\text{gr-out} \right> $ using a moving average with time-window $t_\text{av}$. $D_{80}$ is the $80$\% passing diameter (m), computed from the cumulative size distribution function for mass that enter and leave the mill.
The Bond grindability is then defined as the inverse $G \equiv \text{BWI}^{-1}$. 
We demonstrate the grindability measured in the mill can be backtracked to ore's location of origin for the purpose of improving future estimates for grindability.
Assume there is an a priori, $G_i^0$, estimate of the grindability of the ore in block $i$.
At time $t_k$ a grindability $G(t_k)$ is observed at the mill.
The average mass in the mill originating from the block $i$ is $\left< m_{\text{gr}}^i \right> (t_k)$.
We reconciliate the grindability at block $i$ and time $t_k$ by the weighted average
\begin{equation}\label{eq:grindability_backtracking}
    G_i(t_k) = \frac{\alpha_i^0 G_i^0 + \sum_{k'= 1}^k{\alpha_i(t_{k'}) G(t_{k'})}}{\alpha_i^0 + \sum_{k'= 1}^k{\alpha_i(t_{k'}) }} ,
\end{equation}
where the weight factor is given by the ratio of mass with origin $i$ to the mass in the mill and relative to the block mass $\alpha_i(t_{k'}) = \langle m_{\text{gr}}^i \rangle^2 / m_i^0 \langle m_{\text{gr}} \rangle$. Here, $\langle m_{\text{gr}} \rangle (t_k)$ is the amount of mass in the mill and $m_i^0$ is the total mass of the ore block $i$.
Note that the weight factor is small (or zero) when there is little (or no) mass from the considered block.
The weight factor for the original grindability reflect the confidence of that estimate, $\alpha_i^0 \ll 1$ if the confidence is low and
$\alpha_i^0 \gtrsim 1$ if the confidence is high.
We test backtracking the grindability of the ore originating from the left source for the case of a quadratic hardness profile shown in Fig.~\ref{fig:hardness_vs_MWD_tracking_error}, but mixed in the mill with ore from the right source.
We use a block size of $10$ m and $\alpha^0 = 0.1$.
The BWI signals are shown in Fig.~\ref{fig:grindability} and the grindability backtracked to the source using Eq.~\ref{eq:grindability_backtracking} is found in Fig.~\ref{fig:block_grindability}.

\section{Discussion and conclusion}
A digital twin of a mine may be considered as a system of particle simulations, distributed over virtual assets that evolve according to their physical counterpart using realtime particle simulation where no real sensors or control signals are available.
The simulator tests demonstrate how data analysis and state predictions are made possible by the particle-based material tracking, i.e., in situations where material from different sources are mixed and alter the comminution processes by their natural variations.
Knowing better what material is currently in storage and about to exit enable more efficient milling, by more precise control of the mill speed or by blending ores from different sources.
Understanding how fragmentation and hardness affect the entire chain of processes from loading, crushing to milling, provides the data for deciding what is the most optimal fragmentation from a mine-to-mill perspective.
The realization of particle-based material tracking in a real mine involve creating virtual assets for each of the physical assets, a central historian and tracking service, and interconnecting these in a distributed system.
In a modern mine, most of the components and infrastructure for this is already in place for the purpose of local production planning and optimization.
It is mainly for the storage systems there is a lack of models that provide sufficient resolution and performance for driving the virtual assets.
Particle-based simulation, such as DEM, accelerated using reduced order modeling techniques may fill this need.
The main uncertainties is how sensitive the result is to the resolution of particle size and shape, and to the DEM model parameters including particle mass density, friction etc.
As demonstrated here, this can be investigated in advance using a simulator with material tracking.
Future work should investigate the required level of fidelity of the simulation modes to sufficiently capture how the material disperse over pile surfaces and the velocity field in funnel discharge flow.
Other unknown factors to investigate are the effect of moisture and temperature on the flow properties. 
Calibration and validation tests involving physical tracers is an important tool here.

\vspace{6pt} 

\section*{Supplementary Material}
Supplementary material Video S1: Integration Test and Video S2: Mine Simulator are available at {\small \url{http://umit.cs.umu.se/material_tracking/}}

\section*{Acknowledgements}
The authors are greatful for the support by ABB and Algoryx Simulation in carrying out the demonstration and integration tests, and to Boliden and Epiroc for providing background information about mine-to-mill challenges, MWD data, and the role of material tracking.



\begin{thebibliography}{999}
    \bibitem[Holmberg \em{et~al.}(2017)Holmberg, Kivikytö-Reponen, Härkisaari,
    Valtonen, and Erdemir]{Holmberg2017}
  Holmberg, K.; Kivikytö-Reponen, P.; Härkisaari, P.; Valtonen, K.; Erdemir, A.
  \newblock Global energy consumption due to friction and wear in the mining
    industry.
  \newblock {\em Tribology International} {\bf 2017}, {\em 115},~116--139.
  
  \bibitem[Radziszewski(2013)]{Radziszewski2013}
  Radziszewski, P.
  \newblock Energy recovery potential in comminution processes.
  \newblock {\em Minerals Engineering} {\bf 2013}, {\em 46-47},~83 -- 88.
  
  \bibitem[McKee(2013)]{mckee:2013:umm}
  McKee, I.
  \newblock {\em Understanding Mine to Mill}; The Cooperative Research Centre for
    Optimising Resource Extraction (CRC ORE),  2013.
  
  \bibitem[Grieves(2014)]{grieves2014digital}
  Grieves, M.
  \newblock Digital twin: manufacturing excellence through virtual factory
    replication.
  \newblock {\em White paper} {\bf 2014}, {\em 1},~1--7.
  
  \bibitem[Jeschke \em{et~al.}(2017)Jeschke, Brecher, Meisen, {\"O}zdemir, and
    Eschert]{jeschke2017industrial}
  Jeschke, S.; Brecher, C.; Meisen, T.; {\"O}zdemir, D.; Eschert, T.
  \newblock Industrial internet of things and cyber manufacturing systems. In
    {\em Industrial internet of things}; Springer,  2017; pp. 3--19.
  
  \bibitem[Rossi and Deutsch(2014)]{Rossi2014}
  Rossi, M.; Deutsch, C.
  \newblock {\em Mineral Resource Estimation}; Springer,  2014.
  
  \bibitem[Ouchterlony and Sanchidrián(2019)]{Ouchterlony2019}
  Ouchterlony, F.; Sanchidrián, J.
  \newblock A review of development of better prediction equations for blast
    fragmentation.
  \newblock {\em Journal of Rock Mechanics and Geotechnical Engineering} {\bf
    2019}, {\em 11},~1094 -- 1109.
  
  \bibitem[Evertsson(2000)]{Evertsson2000}
  Evertsson, M.
  \newblock Cone Crusher Performance.
  \newblock PhD thesis, Chalmers University of Technology,,  2000.
  
  \bibitem[Napier‐Munn \em{et~al.}(1996)Napier‐Munn, Morrell, Morrison, and
    Kojovic]{NapierMunn1996}
  Napier‐Munn, T.; Morrell, S.; Morrison, R.; Kojovic, T.
  \newblock {\em Mineral Comminution Circuits Their Operation and Optimisation};
    1996.
  
  \bibitem[Rai \em{et~al.}(2016)Rai, Schunnesson, Lindqvist, and Kumar]{Rai2016}
  Rai, P.; Schunnesson, H.; Lindqvist, P.A.; Kumar, U.
  \newblock Measurement-while-drilling technique and its scope in design and
    prediction of rock blasting.
  \newblock {\em International Journal of Mining Science and Technology} {\bf
    2016}, {\em 26},~711 -- 719.
  
  \bibitem[{Zhou} \em{et~al.}(2011){Zhou}, {Hatherly}, {Ramos}, and
    {Nettleton}]{Zhou2011a}
  {Zhou}, H.; {Hatherly}, P.; {Ramos}, F.; {Nettleton}, E.
  \newblock An adaptive data driven model for characterizing rock properties from
    Drilling data.
  \newblock  2011 IEEE International Conference on Robotics and Automation,
    2011, pp. 1909--1915.
  
  \bibitem[Singh and Narendrula(2006)]{Singh2006}
  Singh, S.P.; Narendrula, R.
  \newblock Factors affecting the productivity of loaders in surface mines.
  \newblock {\em International Journal of Mining, Reclamation and Environment}
    {\bf 2006}, {\em 20},~20--32.
  
  \bibitem[Khorzoughi and Hall(2016)]{Khorzoughi2016}
  Khorzoughi, M.B.; Hall, R.
  \newblock Diggability assessment in open pit mines: a review.
  \newblock {\em International Journal of Mining and Mineral Engineering} {\bf
    2016}, {\em 7},~181--209.
  
  \bibitem[Brunton \em{et~al.}(2003)Brunton, Thornton, Hodson, and
    Sprott]{Brunton2003}
  Brunton, I.D.; Thornton, D.M.; Hodson, R.; Sprott, D.
  \newblock Impact of blast fragmentation on hydraulic excavator dig time.
  \newblock  Proceedings Fifth Large Open Pit Conference,  2003.
  
  \bibitem[Kvarnstr{\"o}m and Bergquist(2012)]{kvarnstrom2012}
  Kvarnstr{\"o}m, B.; Bergquist, B.
  \newblock Improving traceability in continuous processes using flow
    simulations.
  \newblock {\em Production Planning \& Control} {\bf 2012}, {\em 23},~396--404.
  
  \bibitem[Servin \em{et~al.}(2014)Servin, Wang, Lacoursi\'{e}re, and
    Bodin]{Servin2014}
  Servin, M.; Wang, D.; Lacoursi\'{e}re, C.; Bodin, K.
  \newblock Examining the smooth and nonsmooth discrete element approaches to
    granular matter.
  \newblock {\em Int. J. Numer. Meth. Eng.} {\bf 2014}, {\em 97},~878--902.
  
  \bibitem[Wallin and Servin(2021)]{wallin:2021:ddm}
  Wallin, E.; Servin, M.
  \newblock Data-driven model order reduction for granular media.
  \newblock {\em Computational Particle Mechanics} {\bf 2021}.
  \newblock
    doi:{\href{https://doi.org/10.1007/s40571-020-00387-6}{\detokenize{10.1007/s40571-020-00387-6}}}.
  
  \bibitem[Erkayaoğlu(2015)]{Erkayaoglu2015}
  Erkayaoğlu, M.
  \newblock A Data Driven Mine-To-Mill Framework For Modern Mines.
  \newblock PhD thesis,  2015.
  
  \bibitem[Erkayaoglu and Dessureault(2019)]{Erkayaoglu2019}
  Erkayaoglu, M.; Dessureault, S.
  \newblock Improving mine-to-mill by data warehousing and data mining.
  \newblock {\em International Journal of Mining, Reclamation and Environment}
    {\bf 2019}, {\em 33},~409--424.
  
  \bibitem[Benndorf and Buxton(2016)]{Benndorf2016}
  Benndorf, J.; Buxton, M.W.N.
  \newblock Sensor-based real-time resource model reconciliation for improved
    mine production control – a conceptual framework.
  \newblock {\em Mining Technology} {\bf 2016}, {\em 125},~54--64.
  
  \bibitem[{Innes} \em{et~al.}(2011){Innes}, {Nettleton}, and
    {Melkumyan}]{innes:2011:ete}
  {Innes}, C.; {Nettleton}, E.; {Melkumyan}, A.
  \newblock Estimation and tracking of excavated material in mining.
  \newblock  14th International Conference on Information Fusion,  2011, pp.
    1--8.
  
  \bibitem[Innes(2015)]{Innes2015}
  Innes, C.
  \newblock A Stochastic Method for Representation, Modelling and Fusion of
    Excavated Material in Mining.
  \newblock PhD thesis, University of Sydney, 2015.
  
  \bibitem[Wambeke \em{et~al.}(2018)Wambeke, Elder, Miller, Benndorf, and
    Peattie]{Wambeke2018}
  Wambeke, T.; Elder, D.; Miller, A.; Benndorf, J.; Peattie, R.
  \newblock Real-time reconciliation of a geometallurgical model based on ball
    mill performance measurements – a pilot study at the Tropicana gold mine.
  \newblock {\em Mining Technology} {\bf 2018}, {\em 127},~115--130.
  
  \bibitem[S.(2016)]{Zhao2016}
  Zhao, Shi.
  \newblock 3D Real-Time Stockpile Mapping and Modelling with Accurate Quality
    Calculation using Voxels.
  \newblock PhD thesis, University of Adelaide,  2016.
  
  \bibitem[Berton \em{et~al.}(2013)Berton, Jubinville, Hodouin, Prévost, and
    Navarra]{Berton2013}
  Berton, A.; Jubinville, M.; Hodouin, D.; Prévost, C.; Navarra, P.
  \newblock Ore storage simulation for planning a concentrator expansion.
  \newblock {\em Minerals Engineering} {\bf 2013}, {\em 40},~56 -- 66.
  
  \bibitem[ABB()]{ABB800xA}
  {ABB Ability System 800xA}.
  \newblock \url{https://new.abb.com/control-systems/system-800xa}.
  \newblock Accessed: 2021-01-11.
  
  \bibitem[OPC()]{OPC-UA}
  OPC Foundation {Unified Architecture}.
  \newblock \url{https://opcfoundation.org/about/opc-technologies/opc-ua/}.
  \newblock Accessed: 2020-11-02.
  
  \bibitem[AGX()]{AGX21}
  {AGX Dynamics}.
  \newblock \url{https://www.algoryx.se/products/agx-dynamics}.
  \newblock Accessed: 2021-03-29.
  
  \bibitem[Servin \em{et~al.}(2014)Servin, Wang, Lacoursi{\`e}re, and
    Bodin]{servin:2014:esn}
  Servin, M.; Wang, D.; Lacoursi{\`e}re, C.; Bodin, K.
  \newblock Examining the smooth and nonsmooth discrete element approach to
    granular matter.
  \newblock {\em Int. J. Numer. Meth. Engng.} {\bf 2014}, {\em 97},~878--902.
  
  \bibitem[Araker and Bostrom(2020)]{Araker2020}
  Araker, M.; Bostrom, J.
  \newblock Simulation and control of the grinding circuit in Boliden Aitik.
  \newblock Technical Report Technical Report, Boliden Mineral AB and Optimation
    AB,  2020.
  
  \bibitem[Bruchmüller \em{et~al.}(2011)Bruchmüller, {van Wachem}, Gu, and
    Luo]{Bruchmueller2011}
  Bruchmüller, J.; {van Wachem}, B.; Gu, S.; Luo, K.
  \newblock Modelling discrete fragmentation of brittle particles.
  \newblock {\em Powder Technology} {\bf 2011}, {\em 208},~731 -- 739.
  
\end{thebibliography}
\end{document}